\pdfoutput=1
\documentclass[11pt]{article}
\usepackage{amsmath, amsthm, amssymb}    % May not all be necessary
\usepackage{bridges}			% Custom bridges proceedings style
\usepackage{graphicx}			% For including pictures
\usepackage[colorlinks=true, urlcolor=blue, citecolor=black, linkcolor=black]{hyperref}  
\usepackage{subcaption}			% For captioning multi-panel figures

\title{Cutting and Sewing Riemann Surfaces\\
in Mathematics, Physics and Clay}

\author{Nadav Drukker
\vspace{10pt}\\
Department of Mathematics, King's College London,
\\
The Strand, London WC2R 2LS, United Kingdom} % end \author
\date{}					% Suppress any date on submissions

\begin{document}

\maketitle

% Prevent page number 1 from being printed on the first page.
\thispagestyle{empty}

\begin{abstract}
A series of ceramic artworks are presented, inspired by the author's research connecting 
theoretical physics to the beautiful theory of Riemann surfaces. 
More specifically the research is related to 
the classification of curves on the surfaces based on a description of them as 
built from basic building blocks known as ``pairs of pants''.
%\\
The relevant background on this mathematics of these 
two dimensional spaces is outlined, 
some of the artistic process is explained: Both the conceptual ideas and their 
implementation. Many photos of the ceramics are included to illustrate this and 
the connected physics problem is briefly mentioned.
\end{abstract}

\begin{figure}[h!tbp]
	\centering
	\includegraphics[width=2.4in]{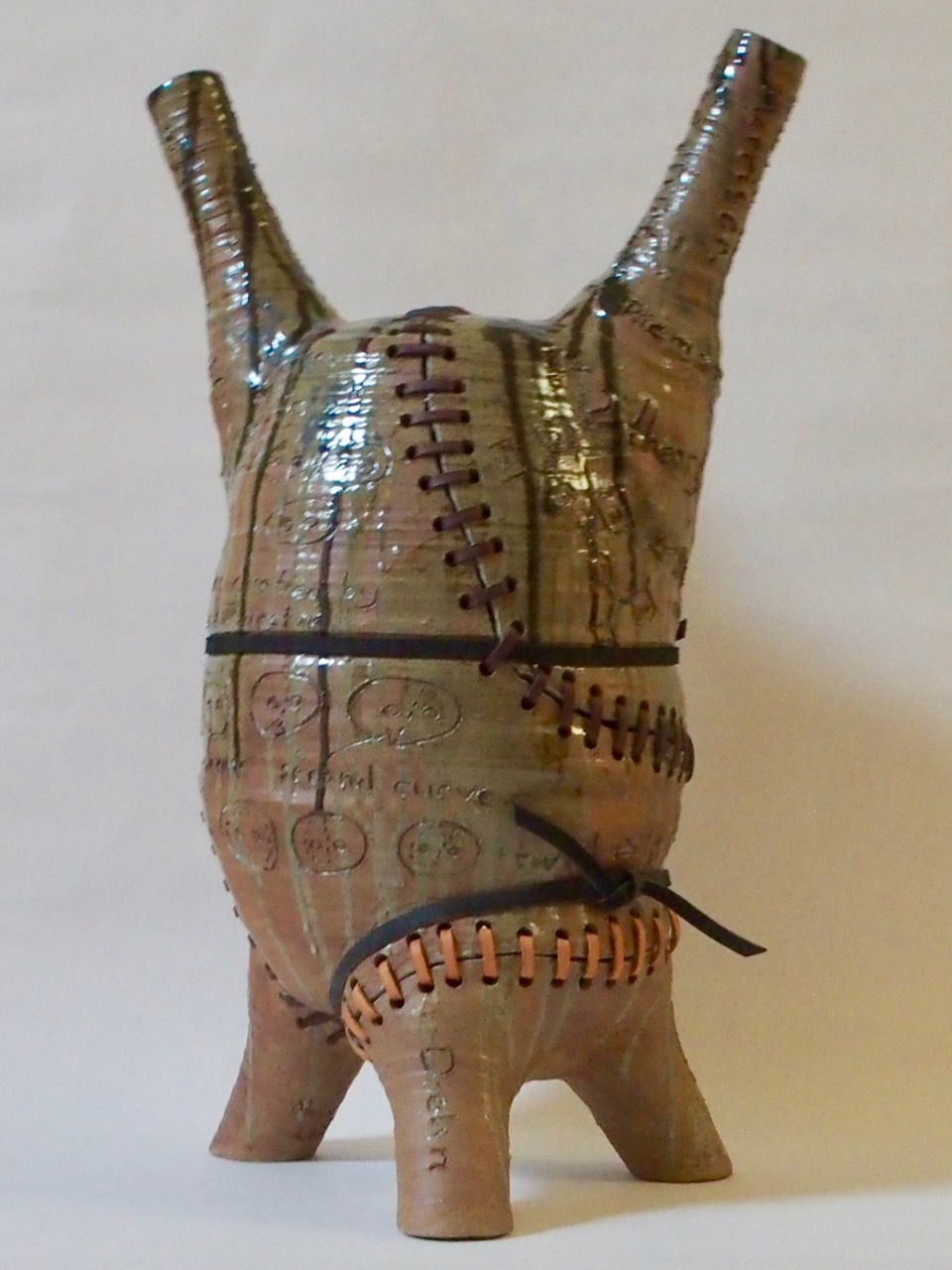}
	\caption{``Sewing-7''. Thrown and assembled stoneware with iron oxide wash, partial 
	glazing and leather straps, $63\times38\times26$cm. The 5-punctured sphere is assembled 
	from three ``pairs of pants''.}
	\label{fig:sewing-7}
\end{figure}

\section*{Introduction}

My speciality as a theoretical physicist is in string theory and supersymmetric 
field theories; both very abstract topics with closer connections to pure mathematics 
than to any observable phenomena. As such it is hard to communicate my research 
to audiences outside my subfield, which is where my hobby of ceramics comes in. 
For every research project that I undertake, I make a series of ceramic vessels or sculptures 
whose forms are inspired by the research and which I inscribe with details of 
the calculations involved.

In this paper I tell the story of a series of artworks titled ``Sewing'', based on my paper 
\cite{DMO} written with T.~Okuda and D.~R.~Morrison. The paper deals with the problem 
of classifying line operators in certain supersymmetric field theories. I explain a bit about 
this question in the last section of this paper and focus mainly on the answer, which is 
easier: The classification exactly matches that of non-self-intersecting 
curves on two dimensional surfaces.

This classification was undertaken by Dehn \cite{Dehn} and independently by 
Thurston \cite{Thurston} (for a detailed discussion, see \cite{PH}) 
and their 
analysis influenced the design of my ceramics. It is based on the pair of pants 
decomposition of surfaces. As I illustrate below, oriented surfaces, possibly with punctures, 
can be described by sewing together basic building blocks which are 
known as pairs of pants (or three-punctured spheres). To be precise, this is true 
for surfaces of negative Euler characteristic and I do not distinguish 
between punctures and finite size holes which is unimportant when viewing the surfaces as topological 
rather than metric or complex spaces. All statements can be refined to the latter cases 
with a bit of care, see \cite{DMO}.

A curve on the surface may be parallel (homotopic) to one of the cuts where it is sewn, 
or it may intersect these cuts, extending from one pair of pants to another. The classification 
of curves (up to topological equivalence, that is smoothly deforming the contour), is then 
given by a set of integers representing the crossings and possible twists. The Dehn-Thurston 
parameters are a pair of integers for each ``cut'' $i$. One, $p_i$, representing the number of crossings and the 
other $q_i$ the overall twist. $p_i\geq0$ and if $p_i=0$, then $q_i$ represents the number of times 
the curve goes around the cut and in this case $q_i\geq0$. In addition, for each pair of pants, the 
sum of the three $p_i$ associated to the three legs has to be even, as we assume the curves are closed.

This classification (presented here very roughly) is quite intuitive and will hopefully become more so 
after reading the examples below. My purpose here is not to give a full account of this topic 
or the physics that is equivalent to it, but rather motivate the ceramic sculptures inspired by them.

\section*{The Once-Punctured Torus}

\begin{figure}[h!tbp]
\centering
\begin{minipage}[b]{0.18\textwidth} 
	\includegraphics[height=4.7cm]{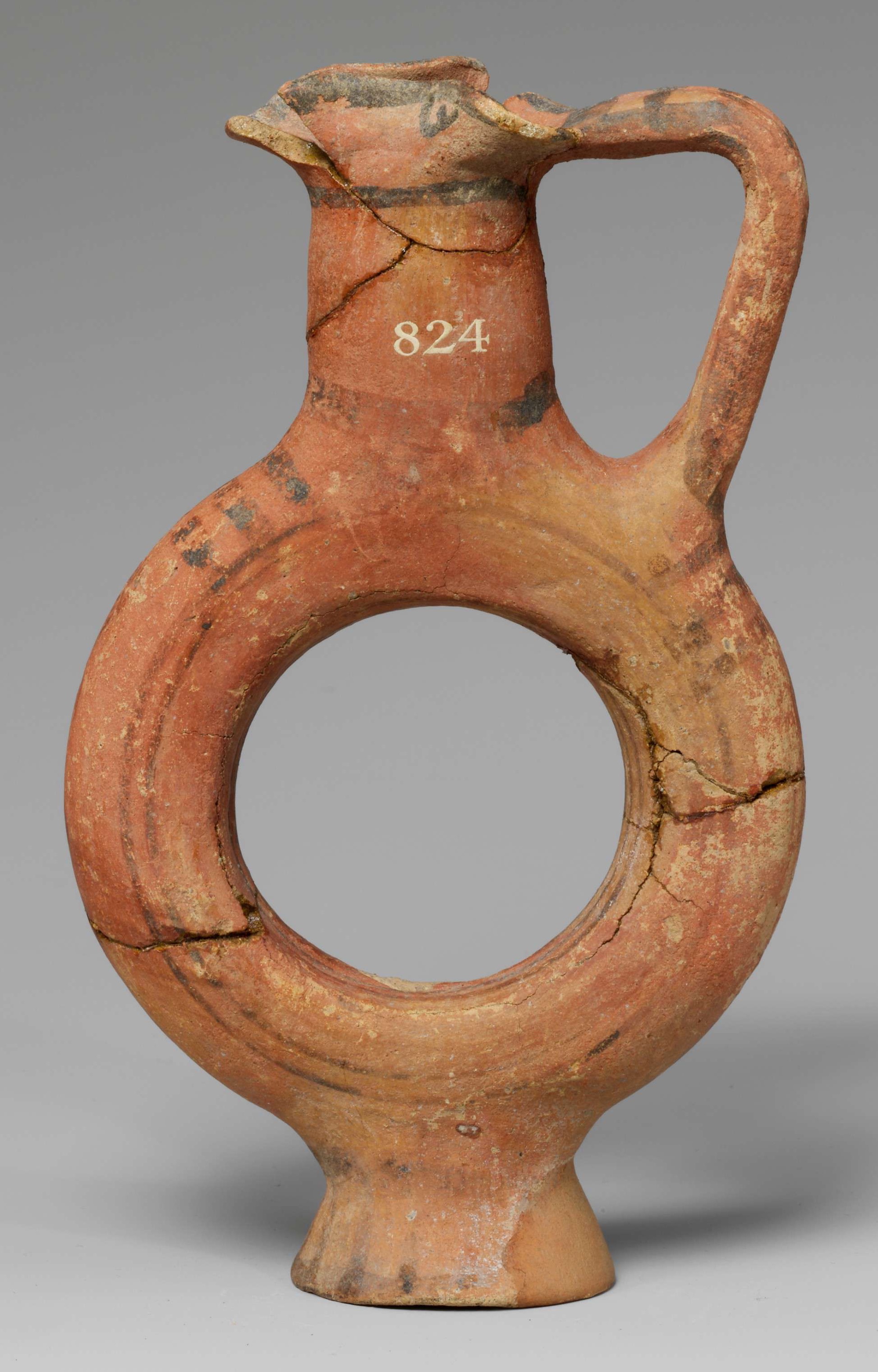}
        	\subcaption*{Cypriot, 850-600B.C., h:$17.1$cm,\\ the Met Museum \cite{met}.}
%        	\label{fig:2c}
\end{minipage}
~ %add desired spacing between images, e. g., ~, \quad, \qquad, \hfill etc.	
\begin{minipage}[b]{0.25\textwidth} 
	\includegraphics[height=4.7cm]{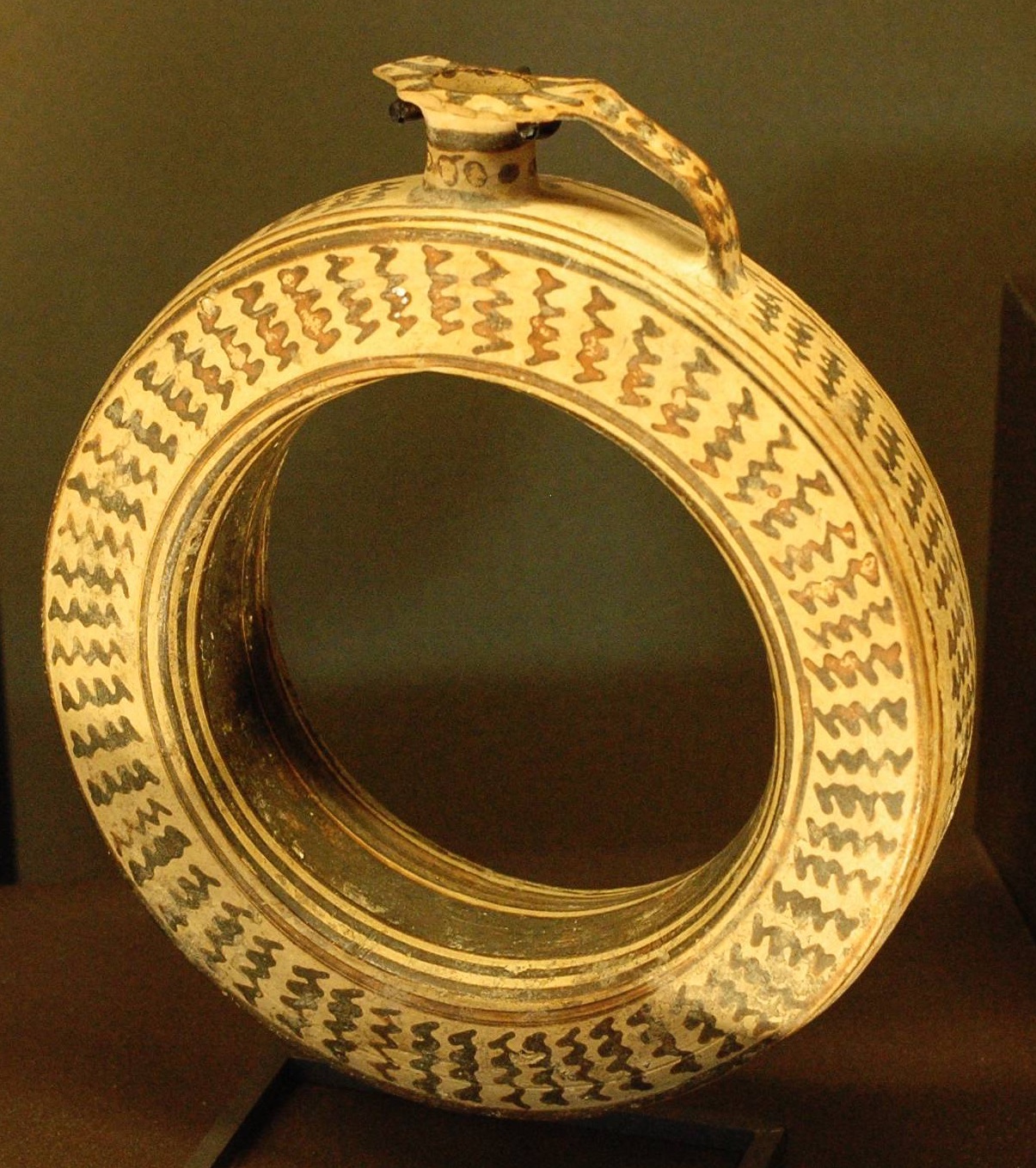}
        	\subcaption*{Protocorinthian, 675-650B.C.,\\ h:$14.7\time13$cm, \\Louvre Museum \cite{louvre}.}
%        	\label{fig:2b}
\end{minipage}
~
\begin{minipage}[b]{0.24\textwidth} 
	\includegraphics[height=4.7cm]{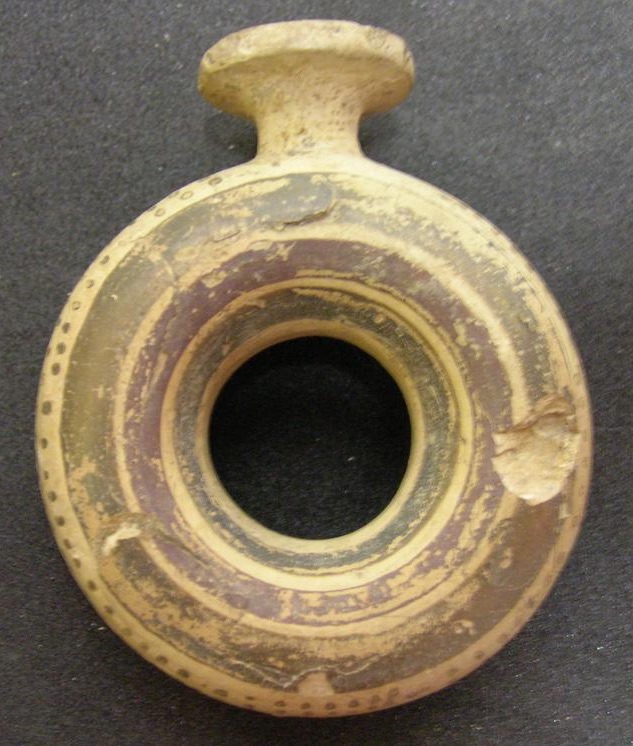}
        	\subcaption*{Early Corinthian, 625-600B.C., $8.2\times6.8\times1.9$cm, \\Ure Museum \cite{reading}.}
%        	\label{fig:2a}
\end{minipage}
~
\begin{minipage}[b]{0.24\textwidth} 
	\includegraphics[height=4.7cm]{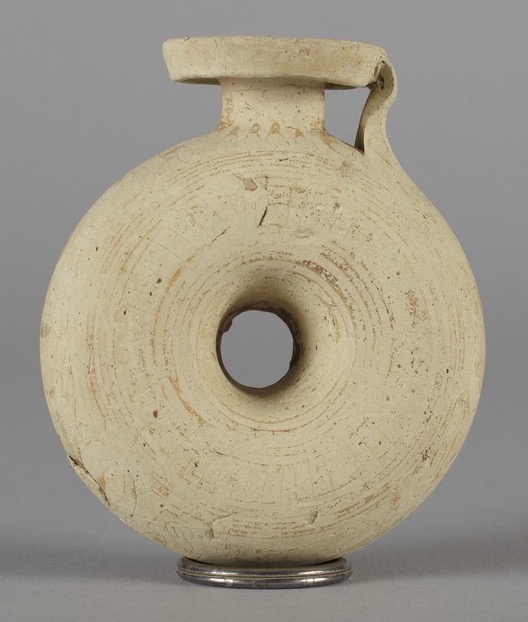}
        	\subcaption*{Late Corinthian, 575-550B.C., $7.4\times6.3\times3.1$cm,\\Princeton Museum \cite{princeton}.}
%        	\label{fig:2c}
\end{minipage}
 %add desired spacing between images, e. g., ~, \quad, \qquad, \hfill etc.	
\caption{Ancient ring aryballi.
%(a) Cypriot, 850-600B.C., height $17.1$cm, Metropolitan Museum of Art \cite{met},\\
%(b) Protocorinthian, 675-650B.C., $14.7\time13$cm, Louvre Museum \cite{louvre},\\
%(b) Early Corinthian, 625-600B.C., $8.2\times6.8\times1.9$cm, Ure Museum \cite{reading},\\
%(a) Late Corinthian, 575-550B.C., $7.4\times6.3\times3.1$cm, Princeton University Museum \cite{princeton}.
}
%\vskip-1cm
\label{fig:old}
\end{figure}

The simplest surfaces studied in \cite{DMO} are the once-punctured torus and the four punctured-sphere. 
The former is made of a single pair 
of pants with two of the legs sewn together, making a torus and leaving one hole. 
The latter requires two pairs of pants sewn together along one of their holes.

\begin{figure}[h!tbp]
\centering
\begin{minipage}[b]{0.2\textwidth} 
	\includegraphics[width=\textwidth]{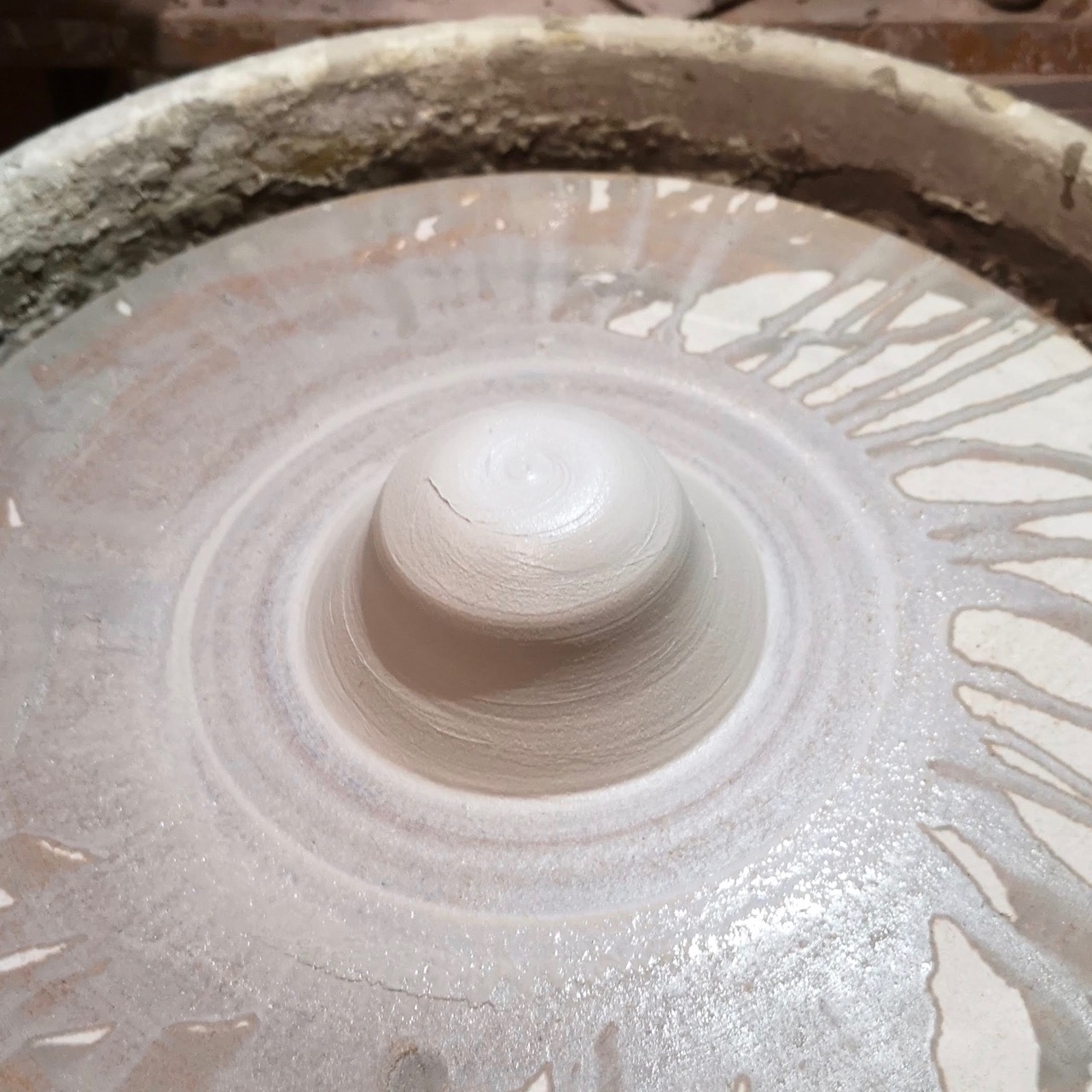}
        	\subcaption*{The clay is centered on the wheel.}
        	\label{fig:process1}
\end{minipage}
~ %add desired spacing between images, e. g., ~, \quad, \qquad, \hfill etc.	
\begin{minipage}[b]{0.2\textwidth} 
	\includegraphics[width=\textwidth]{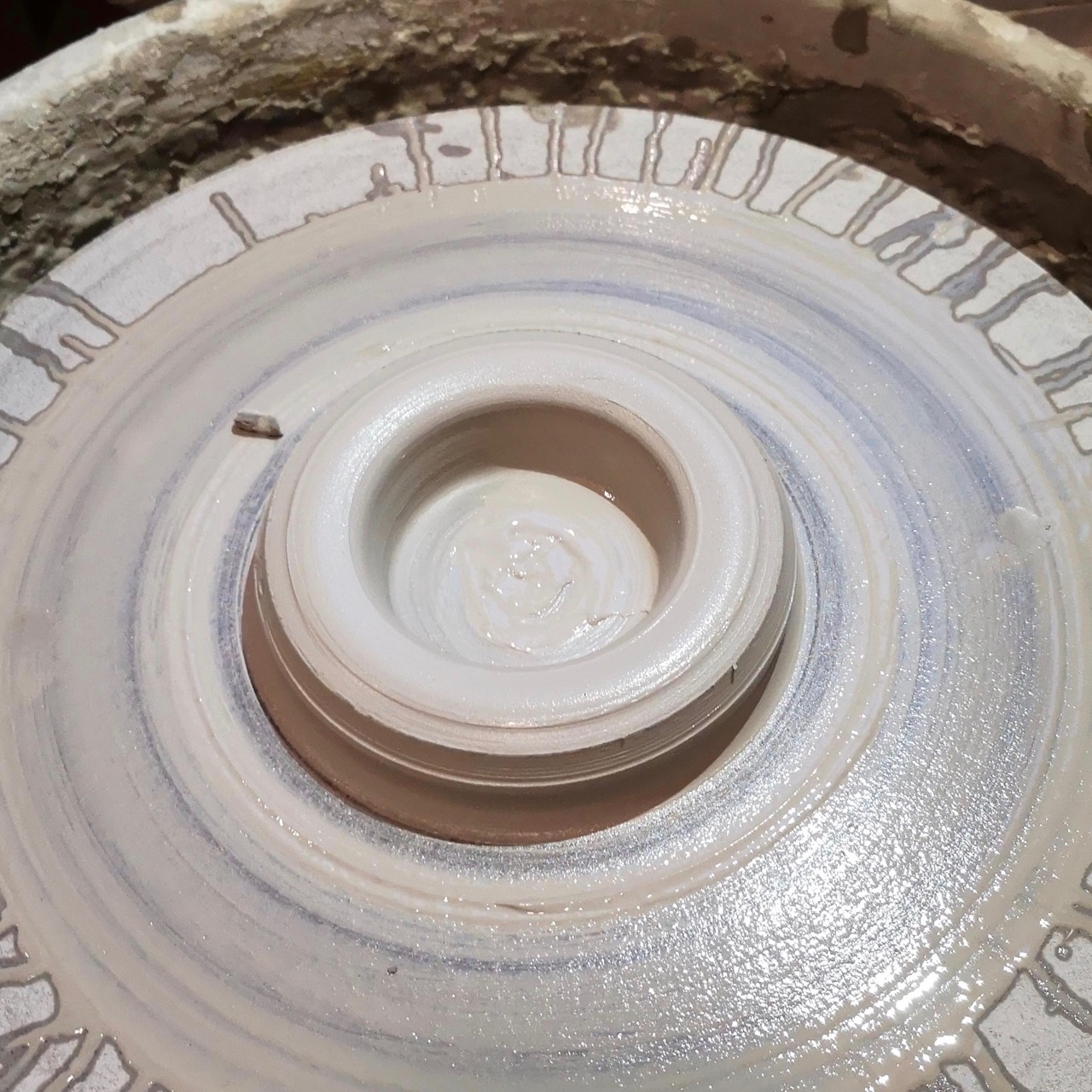}
        	\subcaption*{It is opened up into a solid ring.}
        	\label{fig:process2}
\end{minipage}
~ %add desired spacing between images, e. g., ~, \quad, \qquad, \hfill etc.	
\begin{minipage}[b]{0.2\textwidth} 
	\includegraphics[width=\textwidth]{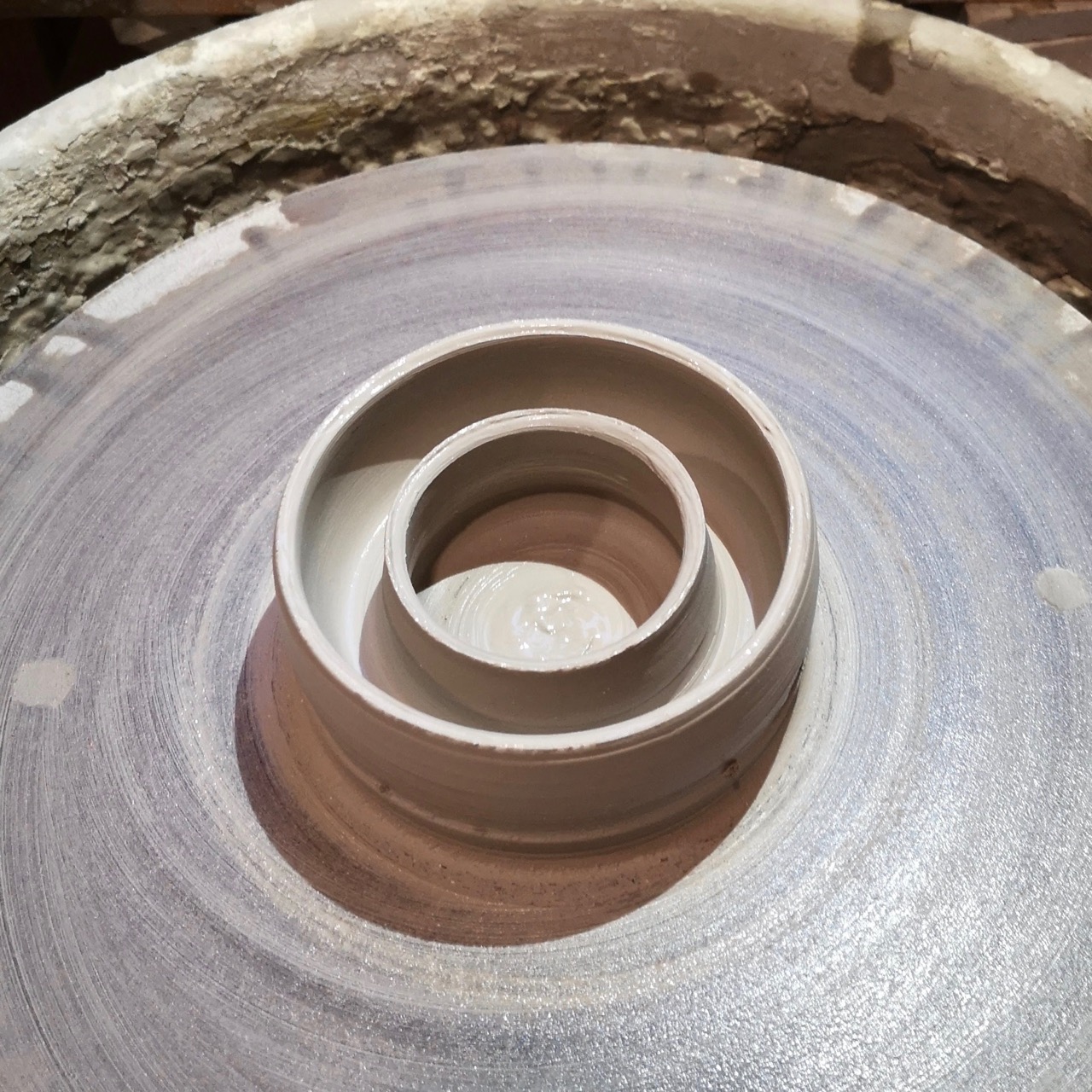}
        	\subcaption*{Two walls are pulled up from the ring.}
        	\label{fig:process3}
\end{minipage}
~ %add desired spacing between images, e. g., ~, \quad, \qquad, \hfill etc.	
\begin{minipage}[b]{0.2\textwidth} 
	\includegraphics[width=\textwidth]{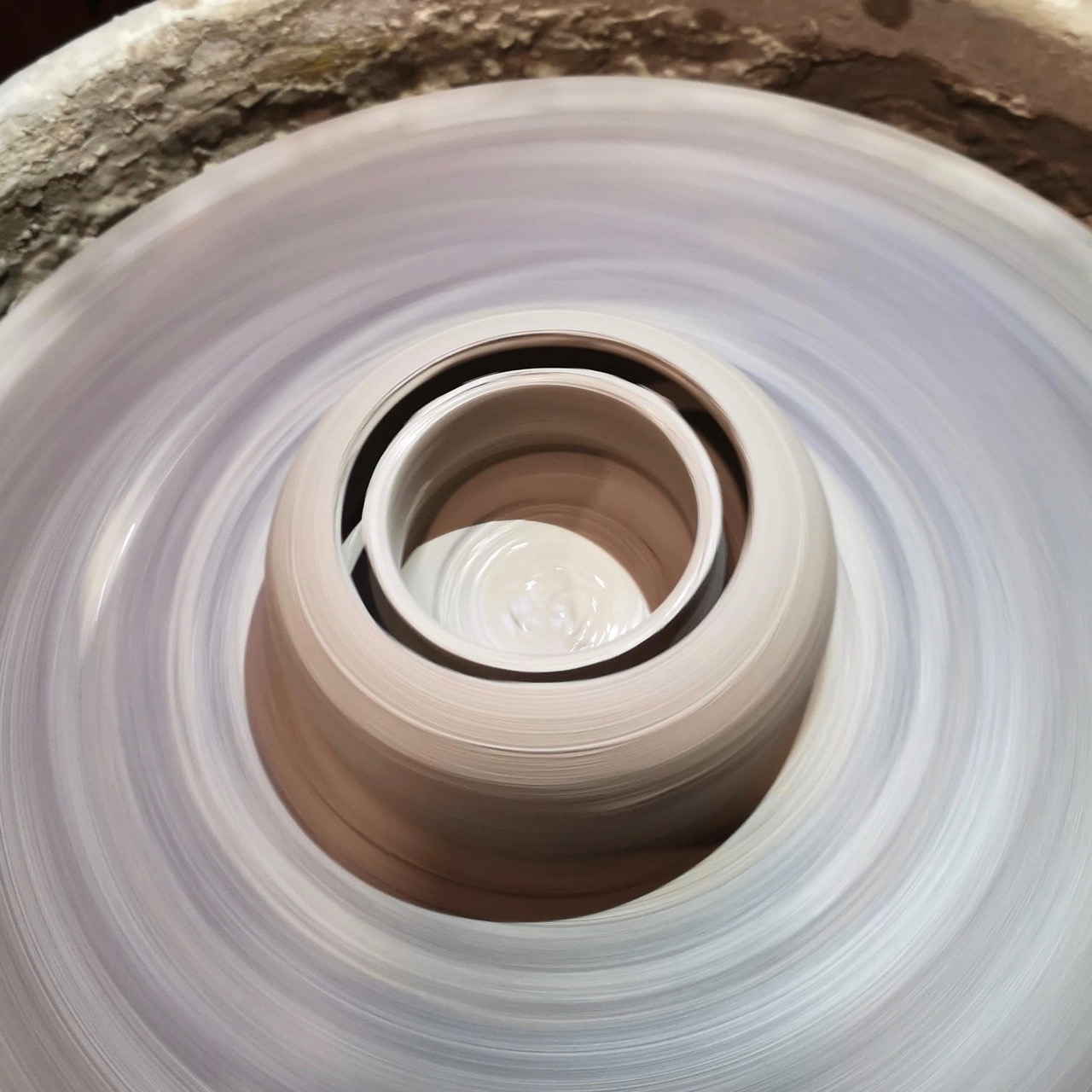}
        	\subcaption*{The walls are bent towards each-other.}
        	\label{fig:process4}
\end{minipage}
\\[2mm]
\begin{minipage}[b]{0.2\textwidth} 
	\includegraphics[width=\textwidth]{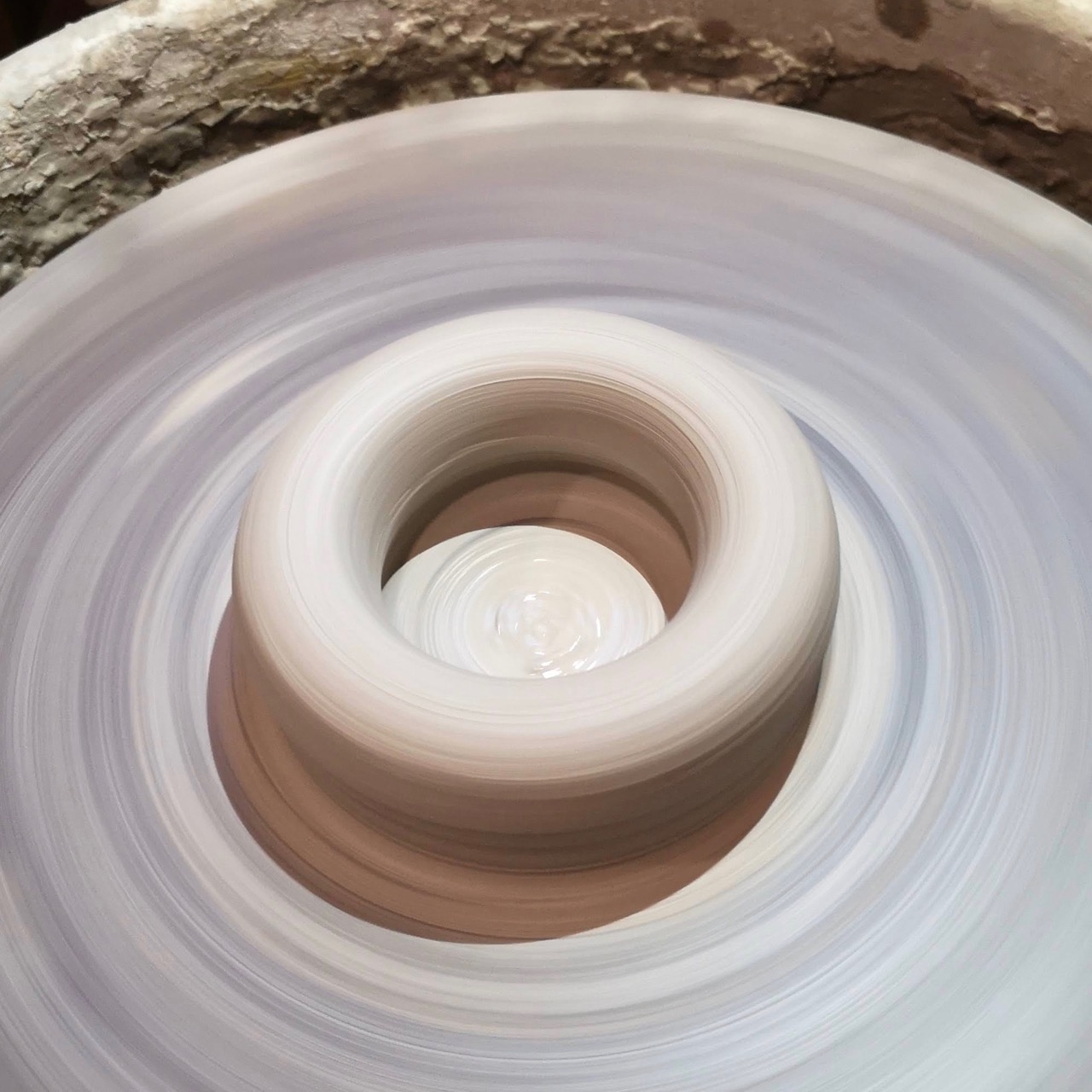}
        	\subcaption*{They are fused to make a hollow torus.}
        	\label{fig:process5}
\end{minipage}
~ %add desired spacing between images, e. g., ~, \quad, \qquad, \hfill etc.	
\begin{minipage}[b]{0.2\textwidth} 
	\includegraphics[width=\textwidth]{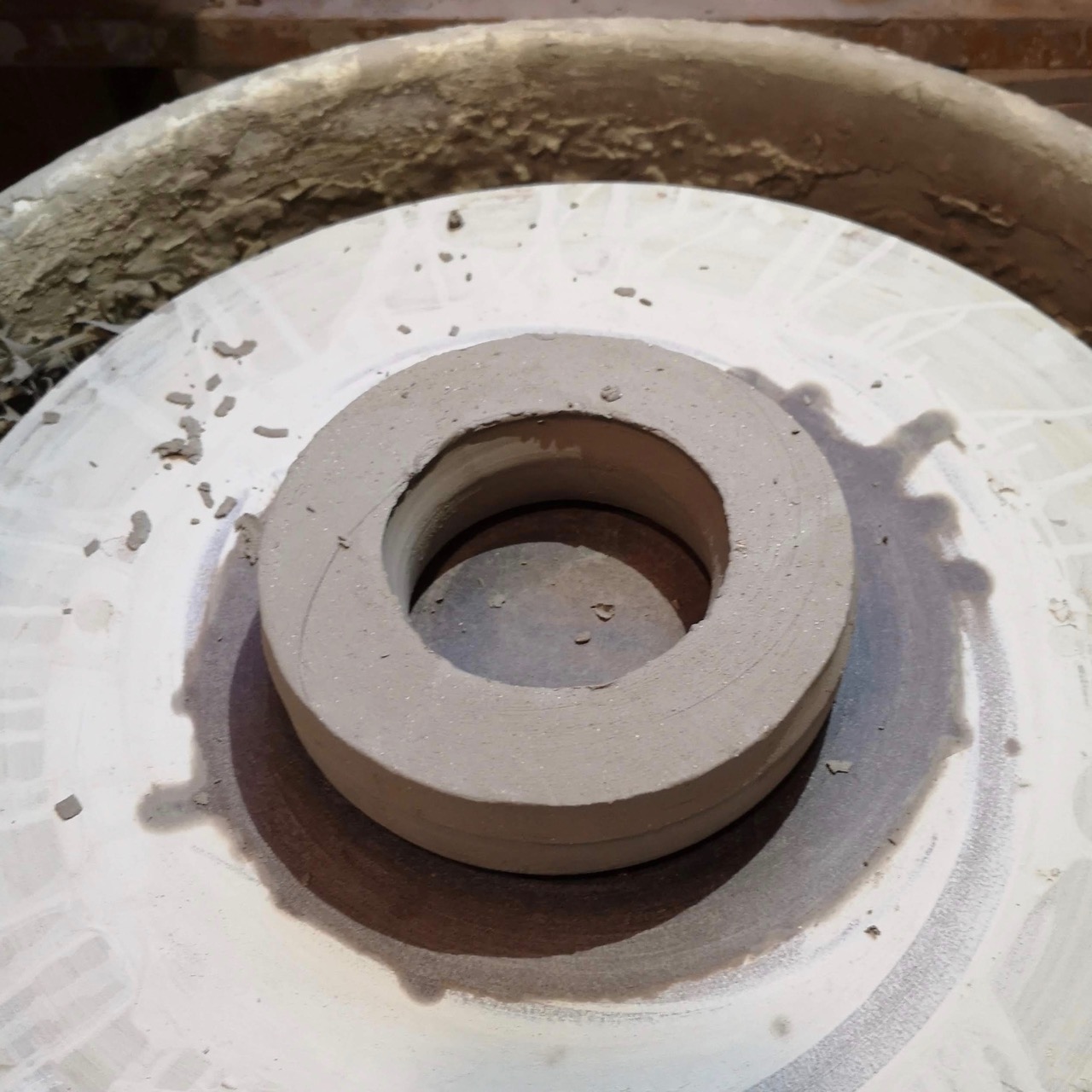}
        	\subcaption*{Partially dried, the ring is placed upside down.}
        	\label{fig:process6}
\end{minipage}
~ %add desired spacing between images, e. g., ~, \quad, \qquad, \hfill etc.	
\begin{minipage}[b]{0.2\textwidth} 
	\includegraphics[width=\textwidth]{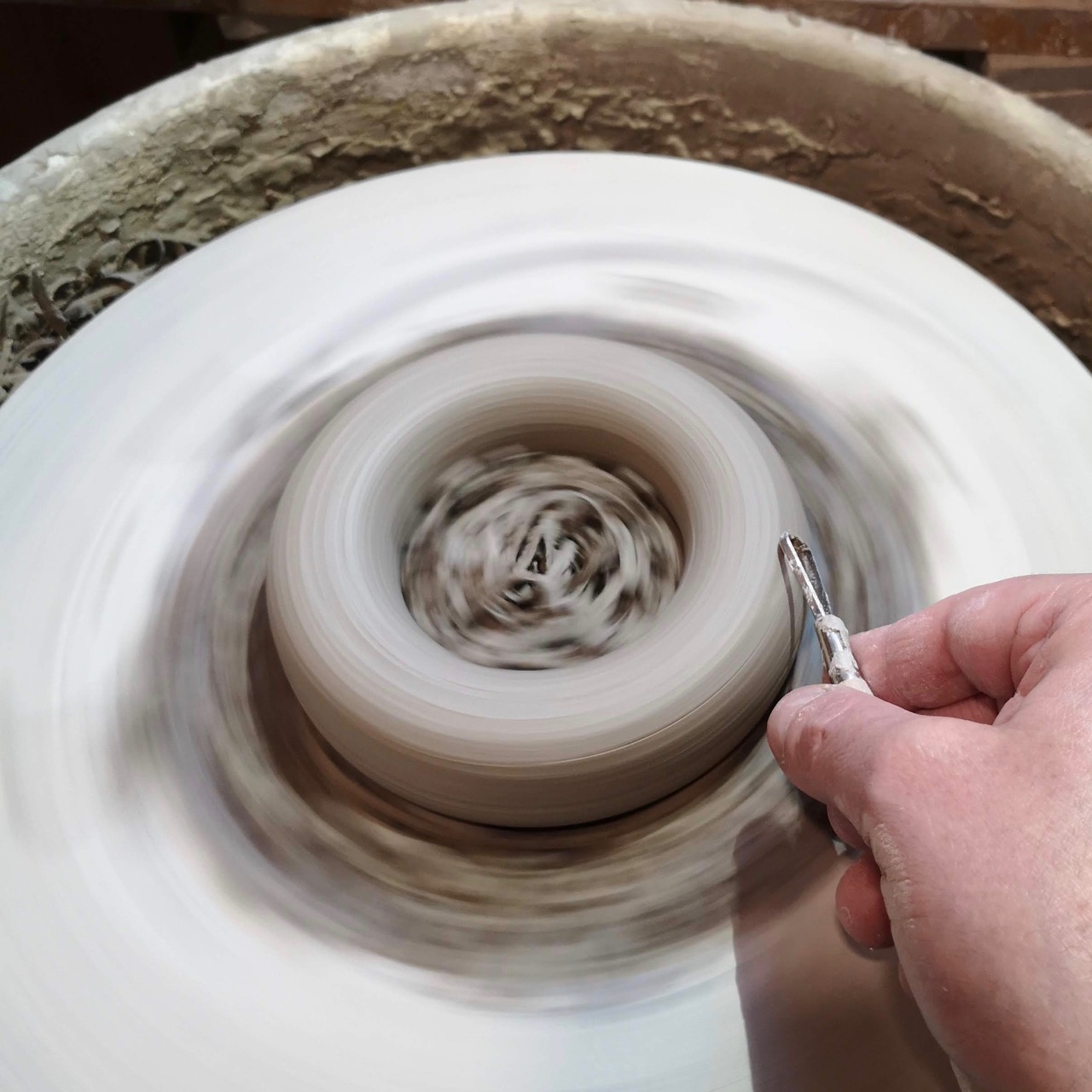}
        	\subcaption*{Excess clay is removed to make it round.}
        	\label{fig:process7}
\end{minipage}
~ %add desired spacing between images, e. g., ~, \quad, \qquad, \hfill etc.	
\begin{minipage}[b]{0.2\textwidth} 
	\includegraphics[width=\textwidth]{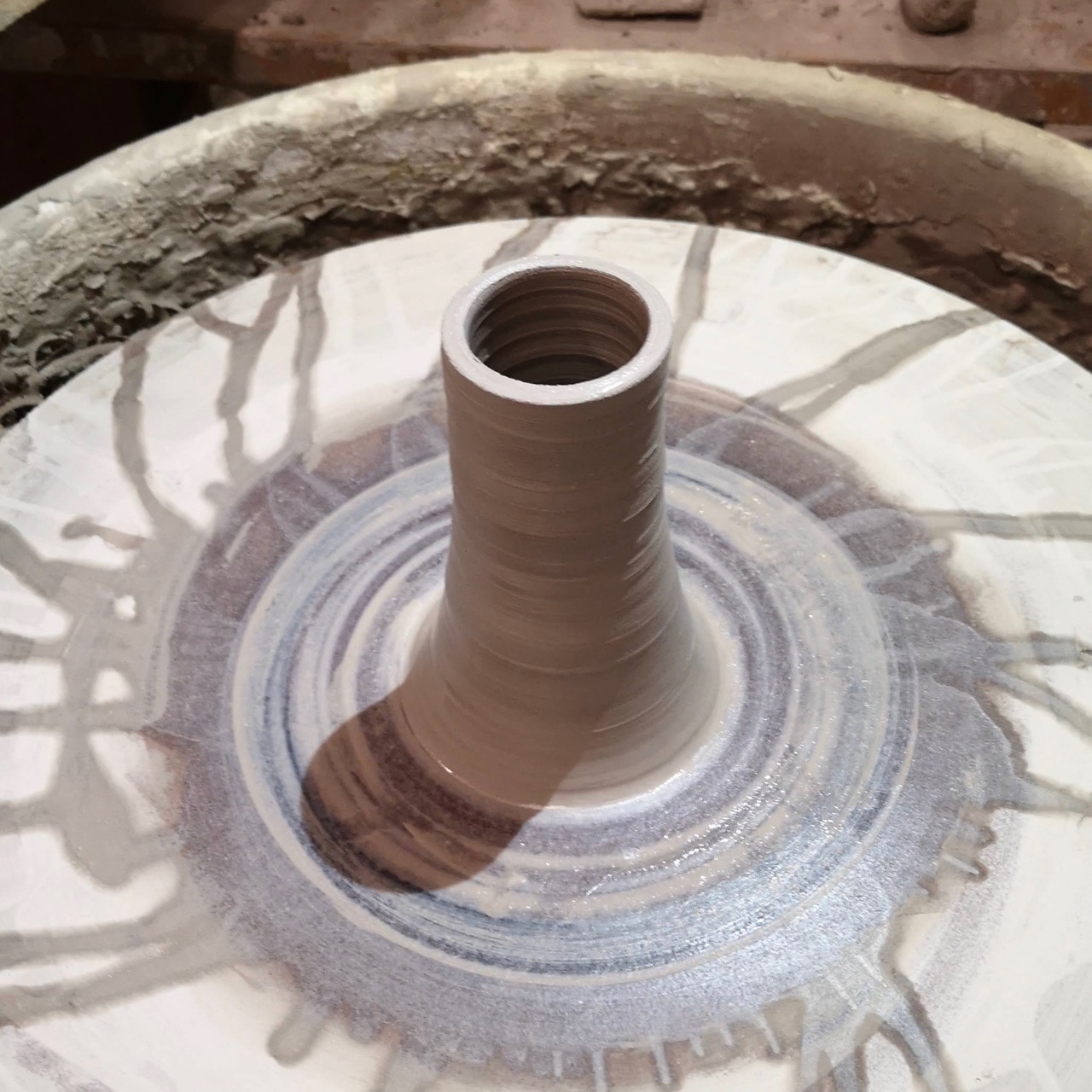}
        	\subcaption*{A tubular base is made and later joined.}
        	\label{fig:process8}
\end{minipage}
\caption{Making a once-punctured torus on the wheel}
\label{fig:process}
\end{figure}

There is a long tradition of of making toroidal jugs \cite{ure}, see Figure~\ref{fig:old} for some early examples. 

\begin{figure}[h!tbp]
\centering
\begin{minipage}[b]{0.22\textwidth} 
	\includegraphics[width=\textwidth]{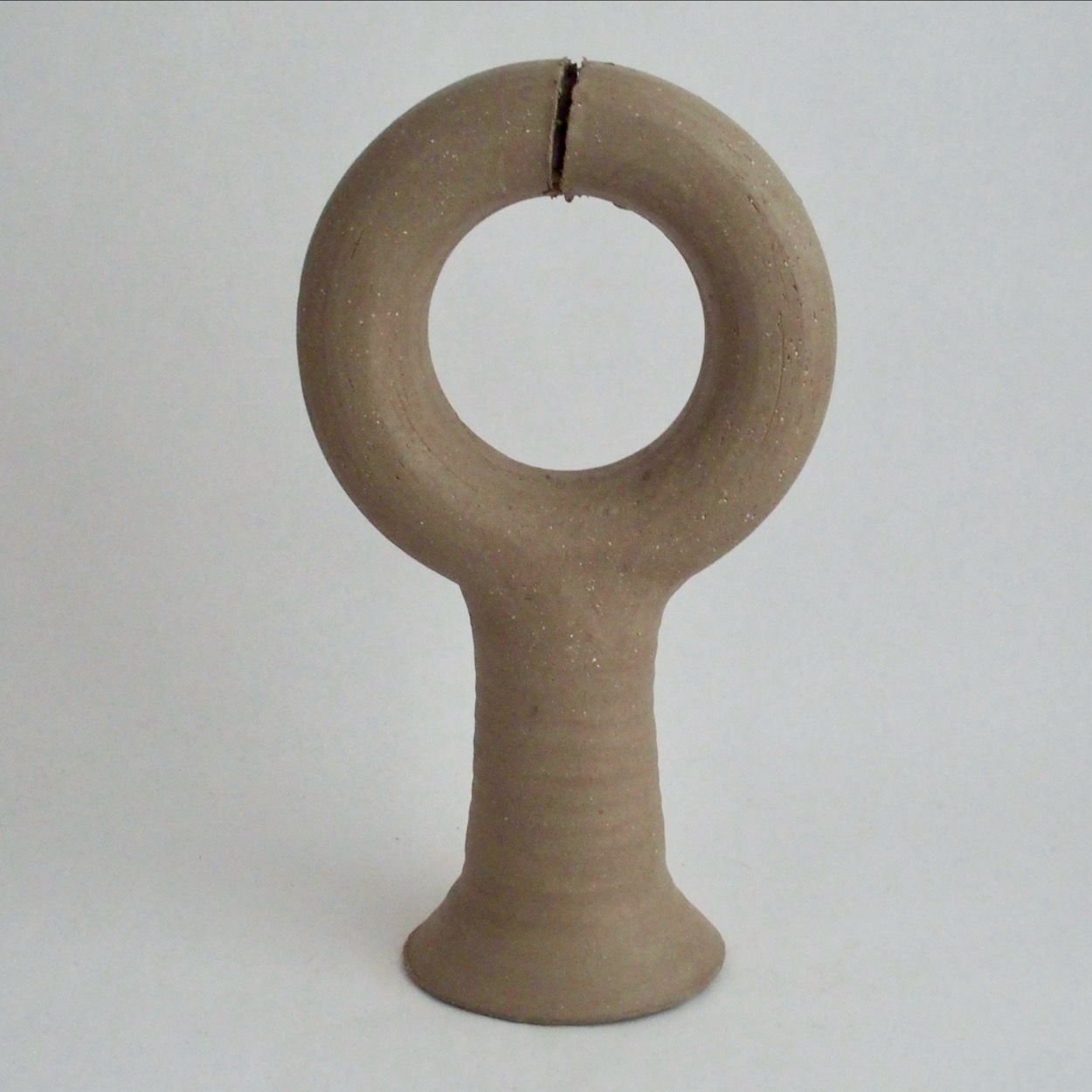}
\end{minipage}
~\begin{minipage}[b]{0.22\textwidth} 
	\includegraphics[width=\textwidth]{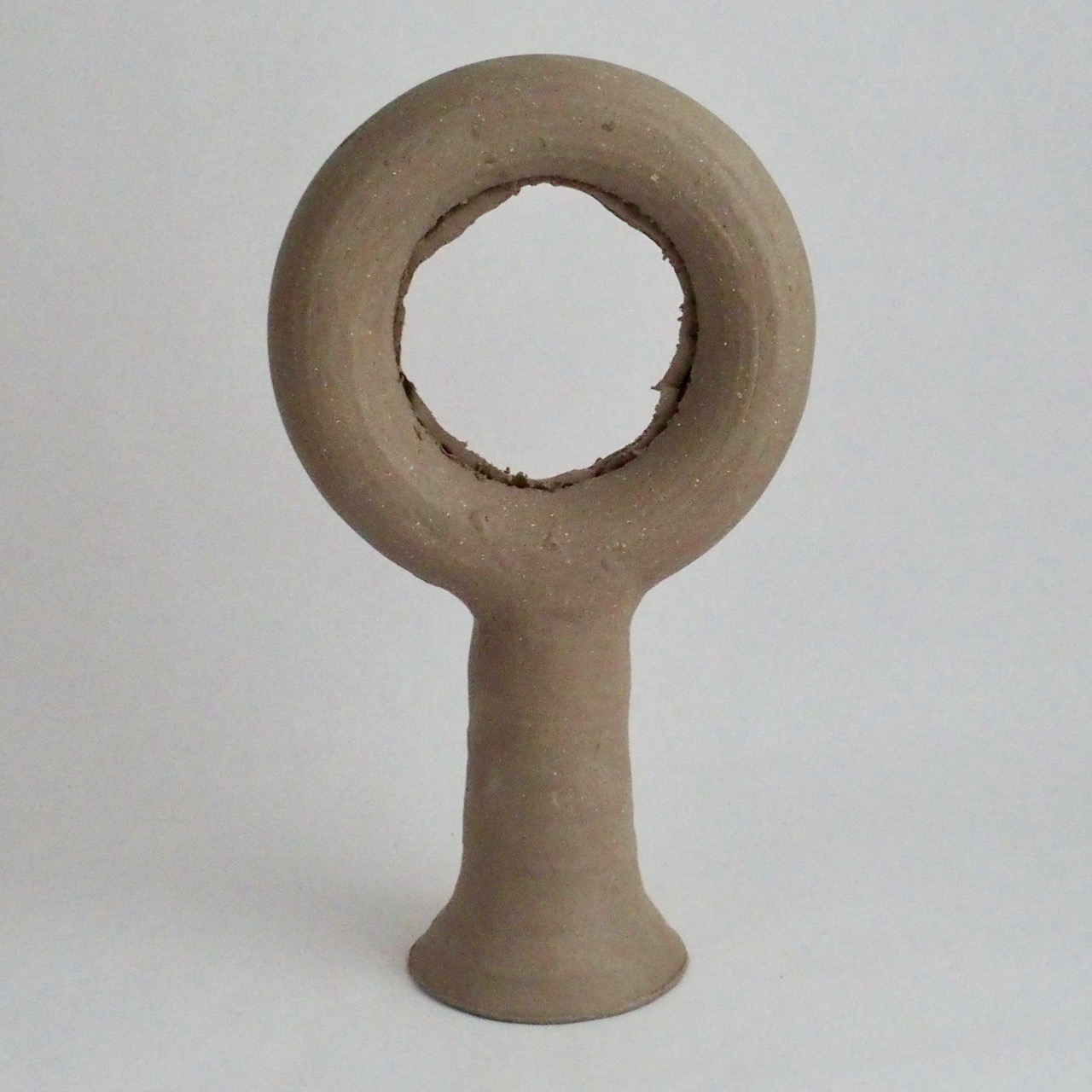}
\end{minipage}
~
\begin{minipage}[b]{0.22\textwidth} 
	\includegraphics[width=\textwidth]{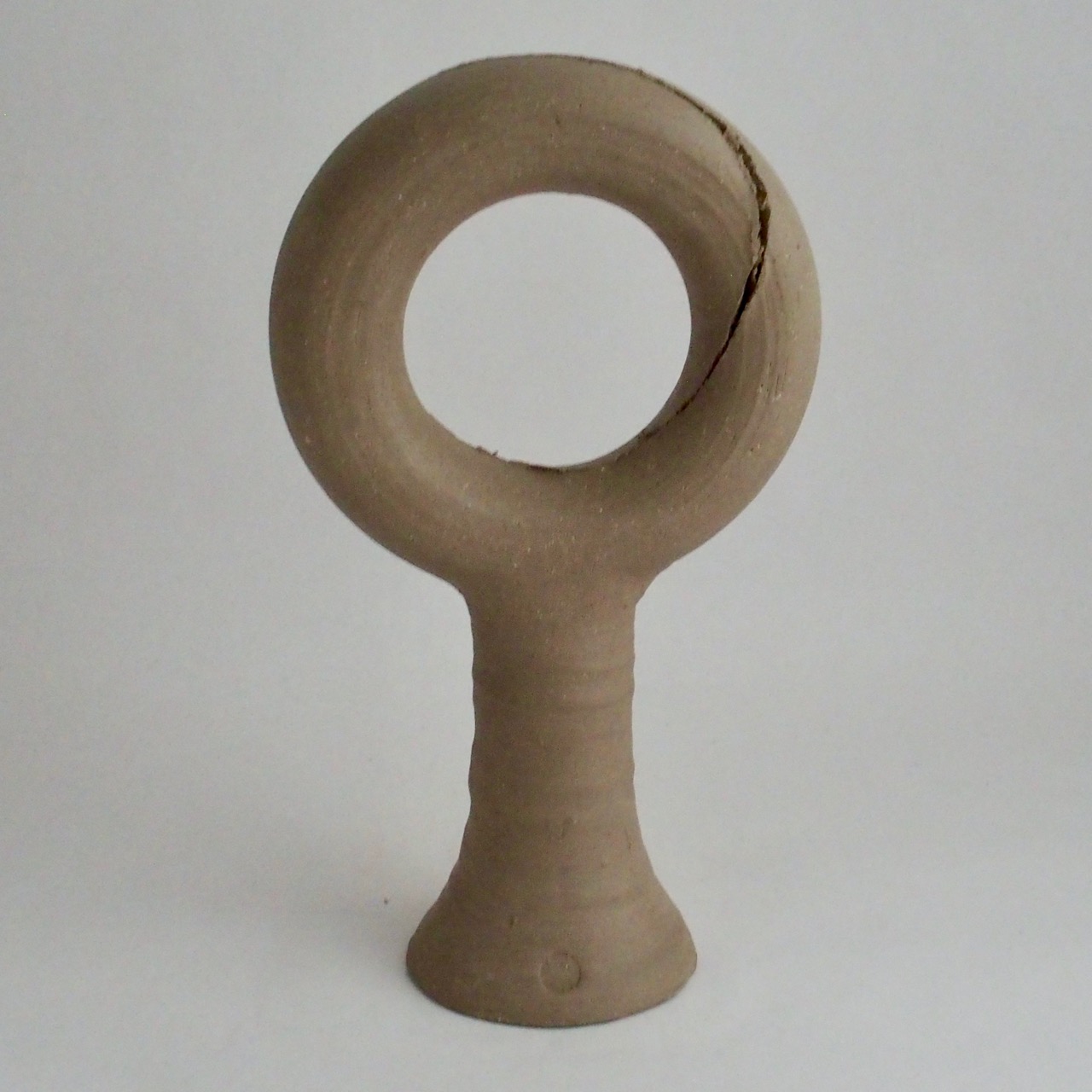}
\end{minipage}
\\%[-3mm]
\begin{minipage}[b]{0.22\textwidth} 
	\includegraphics[width=\textwidth]{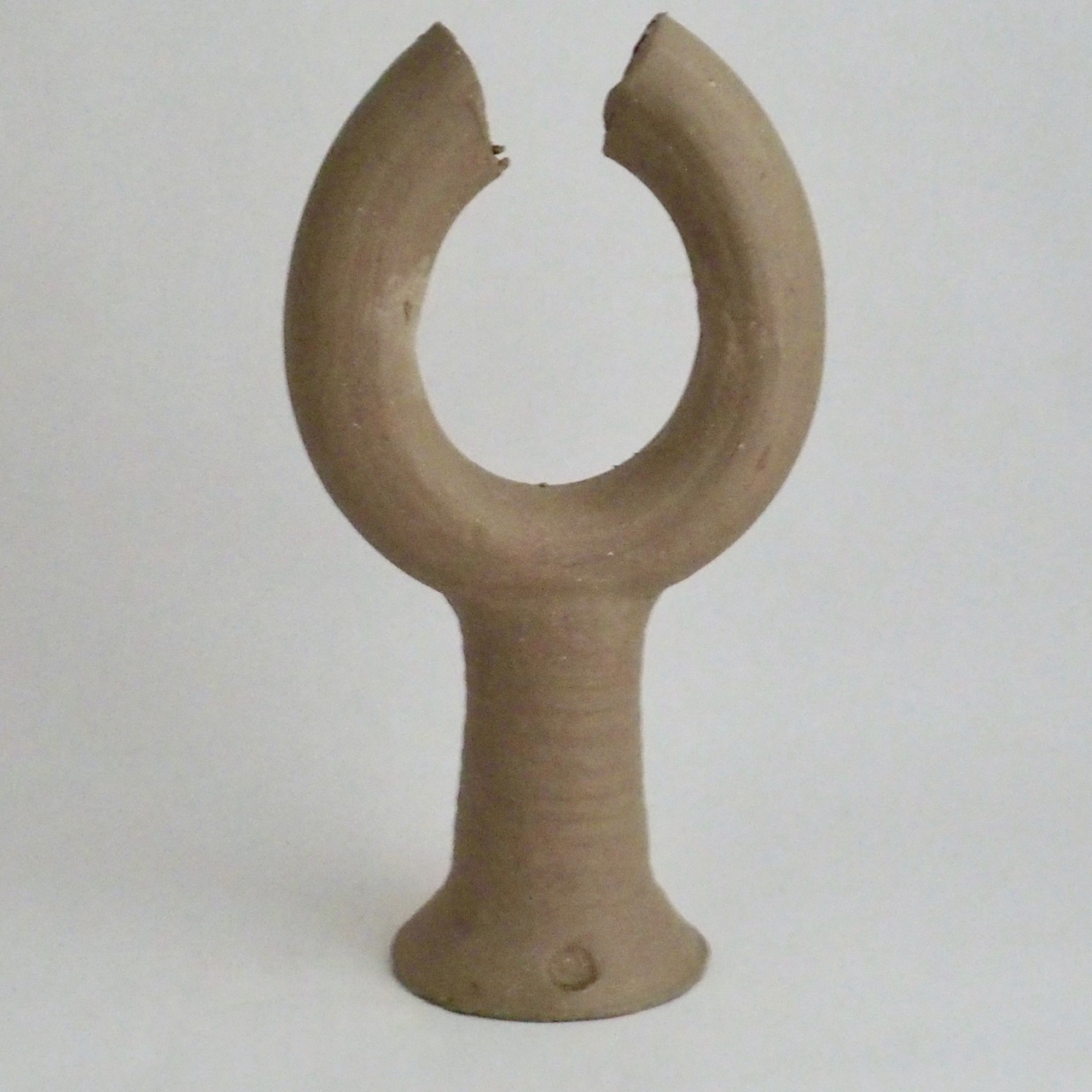}
        	\subcaption*{Cut along the short cycle.}
%        	\label{fig:cutting1}
\end{minipage}
~
\begin{minipage}[b]{0.22\textwidth} 
	\includegraphics[width=\textwidth]{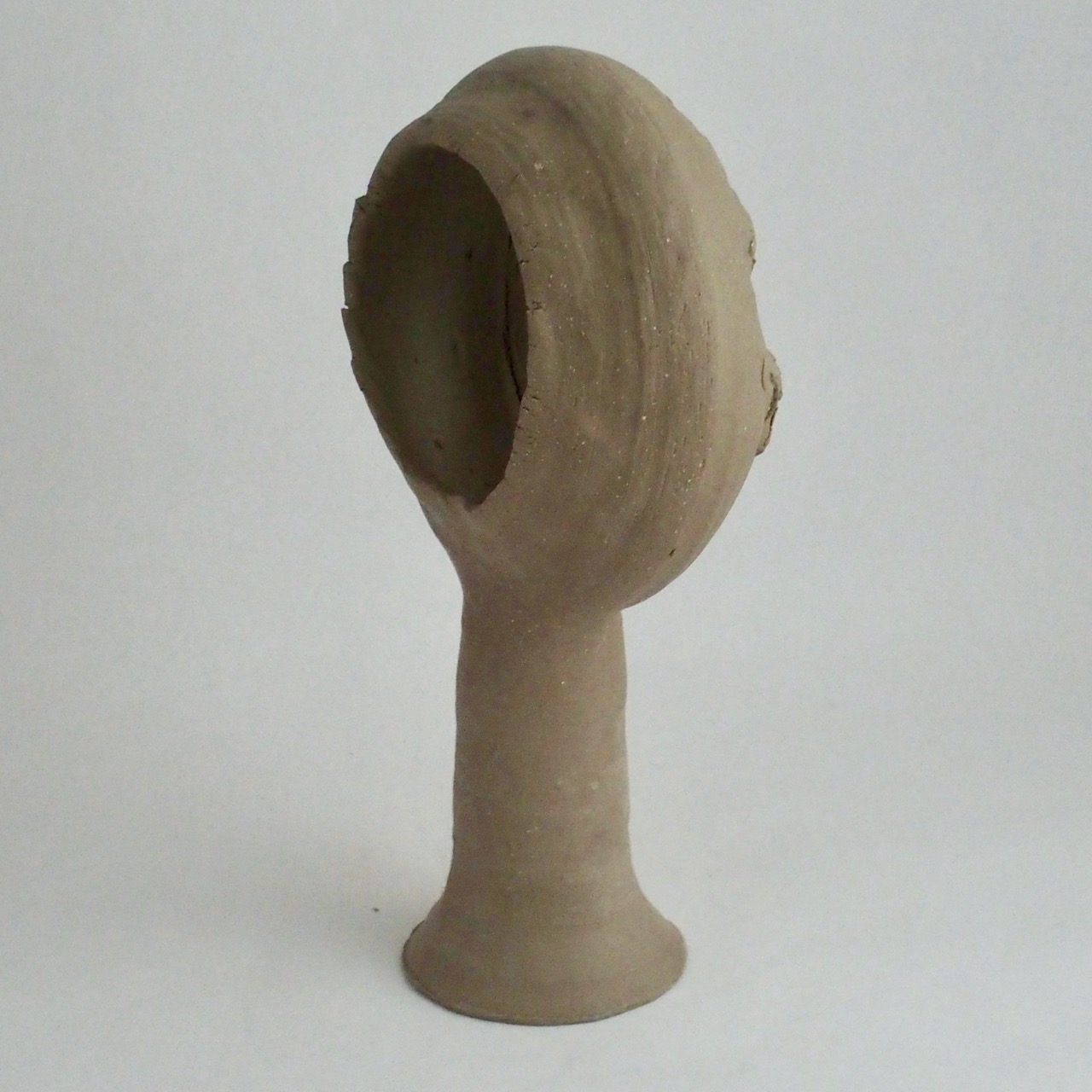}
        	\subcaption*{Cut along the long cycle.}
%        	\label{fig:cutting2}
\end{minipage}
~
\begin{minipage}[b]{0.22\textwidth} 
	\includegraphics[width=\textwidth]{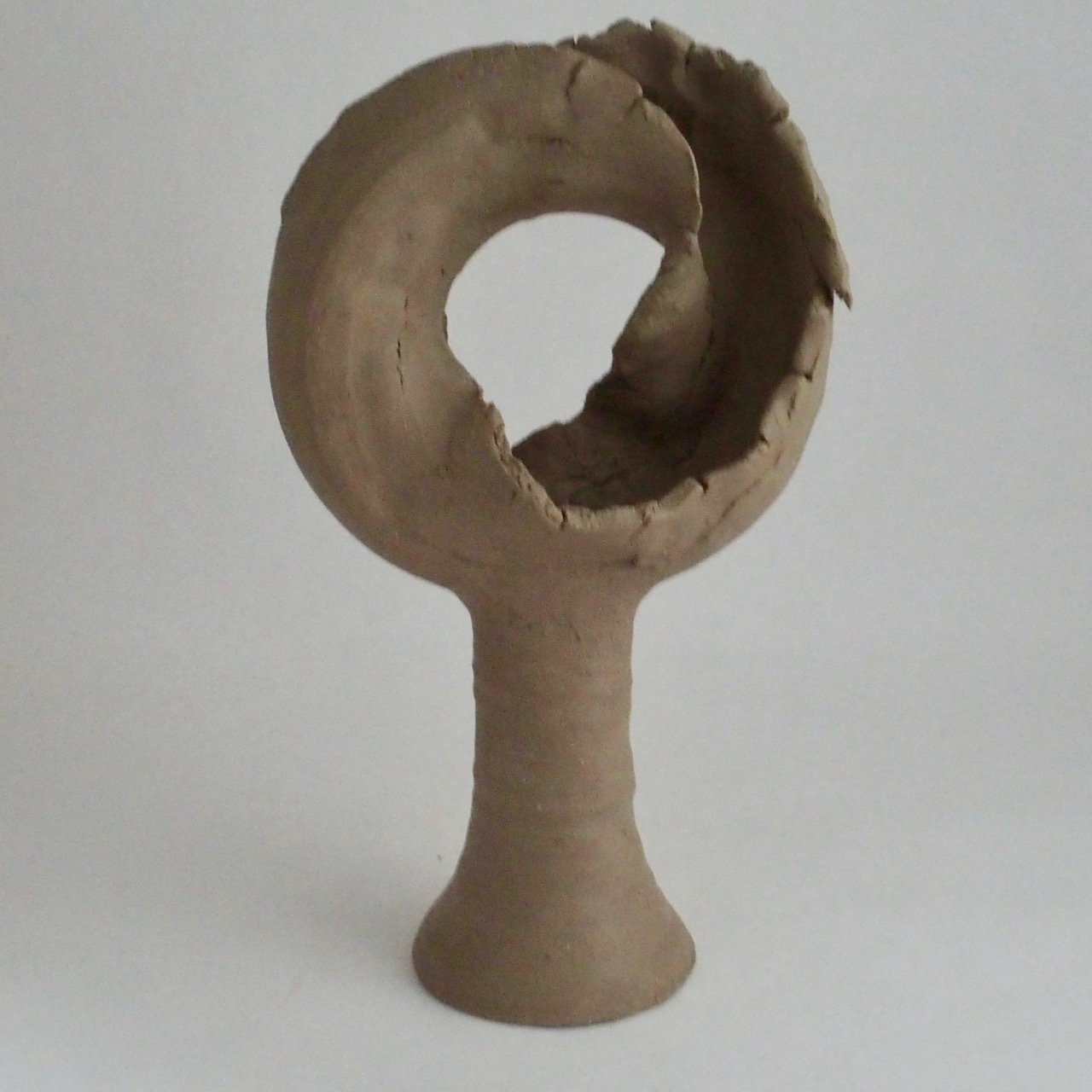}
        	\subcaption*{Cut winding both cycles.}
%        	\label{fig:cutting3}
\end{minipage}
\caption{Cutting and deforming the once-punctured torus.}
\label{fig:cutting}
\end{figure}

Figure~\ref{fig:process} illustrates the steps I use in making a torus on a modern potter's wheel. See the captions for details.
The resulting piece is a torus on a stand where the latter represents a puncture in the torus. 
There in fact has to be a puncture, as one cannot fire ceramic pieces with trapped air. 
As mentioned above, such a surface can be viewed as a single pair of pants with two legs sewn together. 
Conversely, we can cut the punctured torus to get the pair of pants, as I illustrate in Figure~\ref{fig:cutting}.

\noindent
There are an infinite numbers of ways to do that, but I made only three. It is easy to see that 
after the cut we have a surface with three boundaries, which is what is meant by a pair of pants. 
The third example in Figure~\ref{fig:cutting} is a bit counter-intuitive, and does not look like normal trousers. 
rather two of the boundaries are circles which are linked with each other. If we were to 
trace these circles with yarn, the two loops would be interlocked. Still, as far as the intrinsic 
(two dimensional) properties 
of the surface is concerned, it is not really different than the other two examples, 
and this arises because of the way it is embedded in three dimensions.

Other pair of pants decompositions would get more and more complicated---the two circles 
would twist around each other multiple times.

\begin{figure}[h!tbp]
\centering
\begin{minipage}[b]{0.32\textwidth} 
	\includegraphics[width=\textwidth]{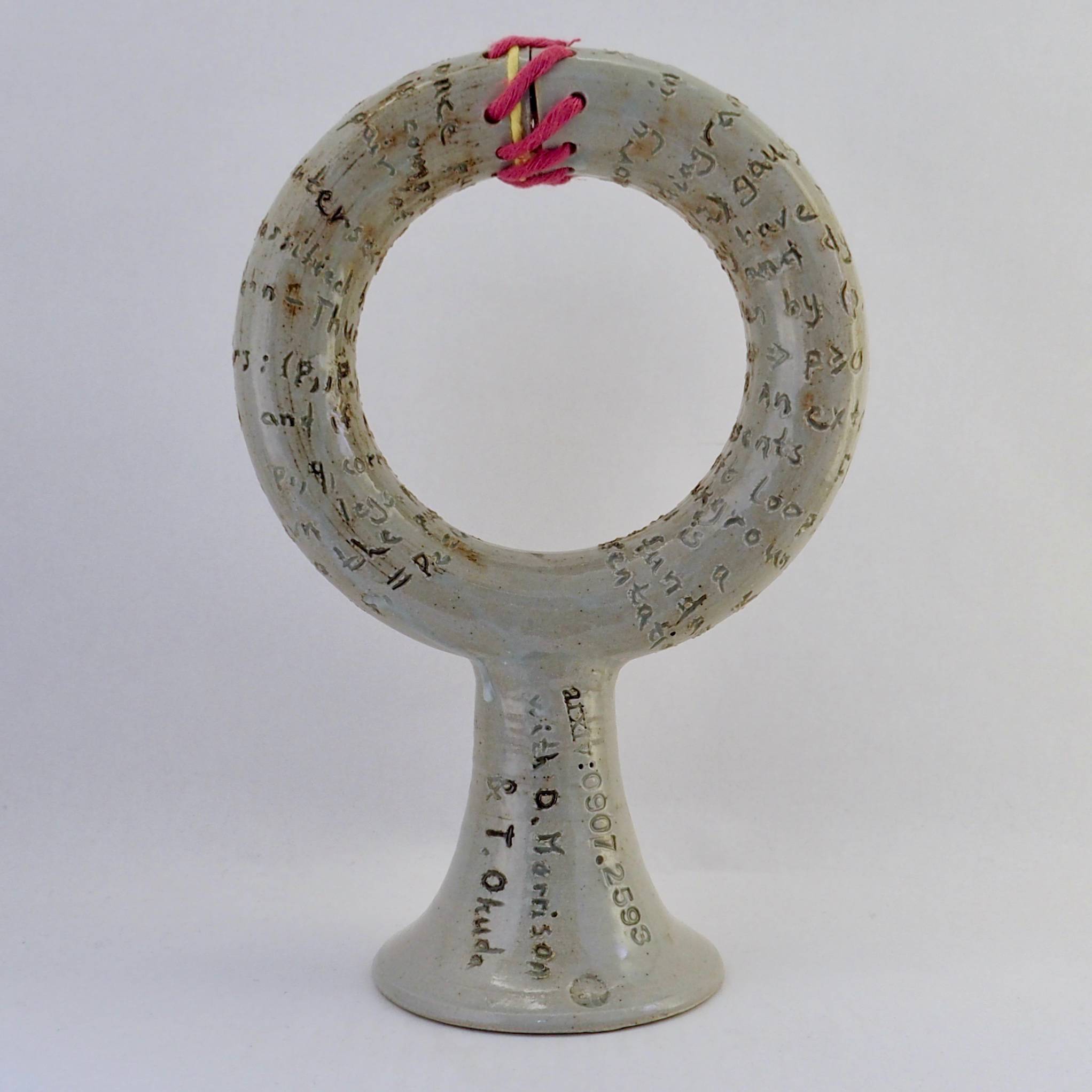}
%        	\subcaption{}
%        	\label{fig:2a}
\end{minipage}
~ %add desired spacing between images, e. g., ~, \quad, \qquad, \hfill etc.	
\begin{minipage}[b]{0.32\textwidth} 
	\includegraphics[width=\textwidth]{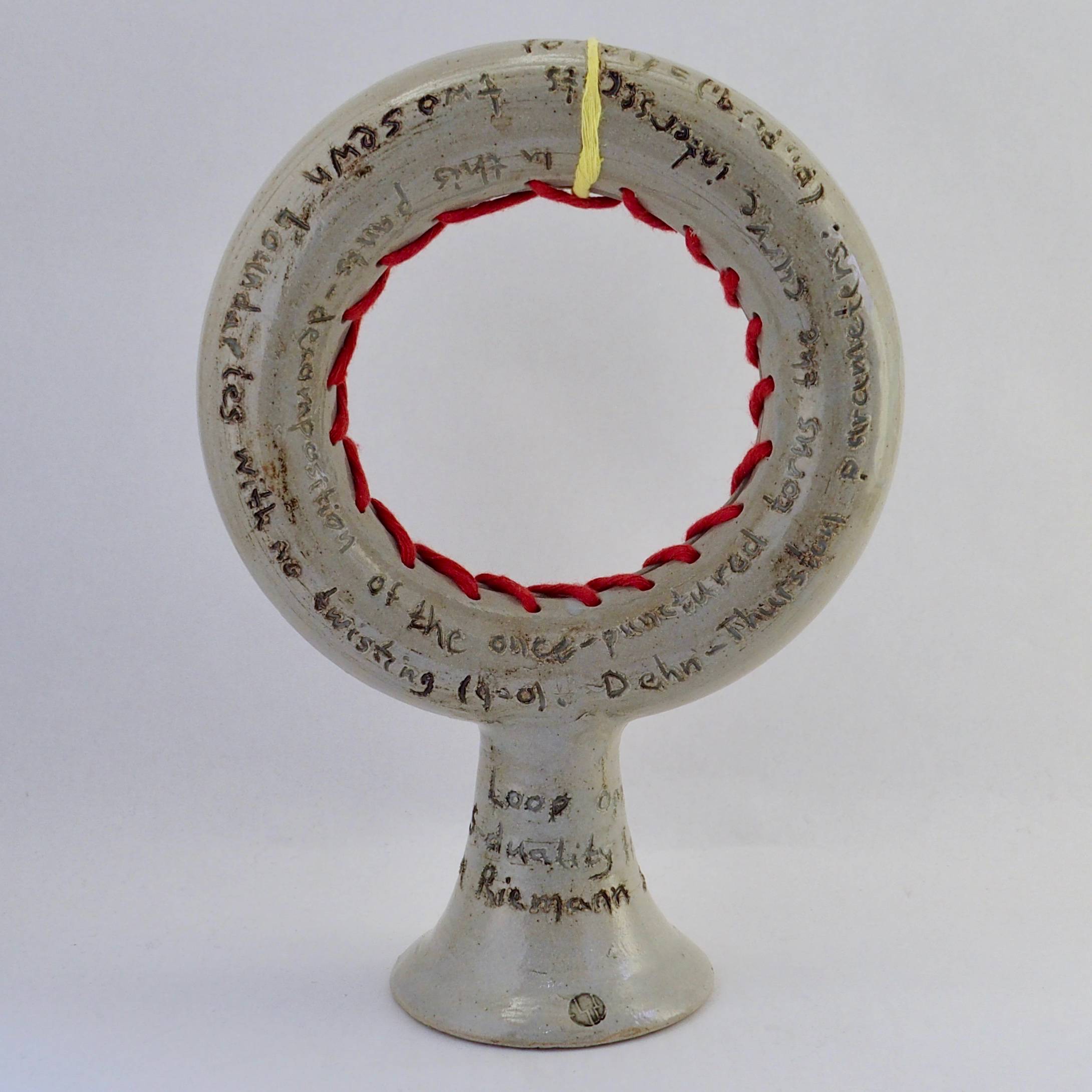}
%        	\subcaption{}
%        	\label{fig:2b}
\end{minipage}
~ %add desired spacing between images, e. g., ~, \quad, \qquad, \hfill etc.	
\begin{minipage}[b]{0.32\textwidth} 
	\includegraphics[width=\textwidth]{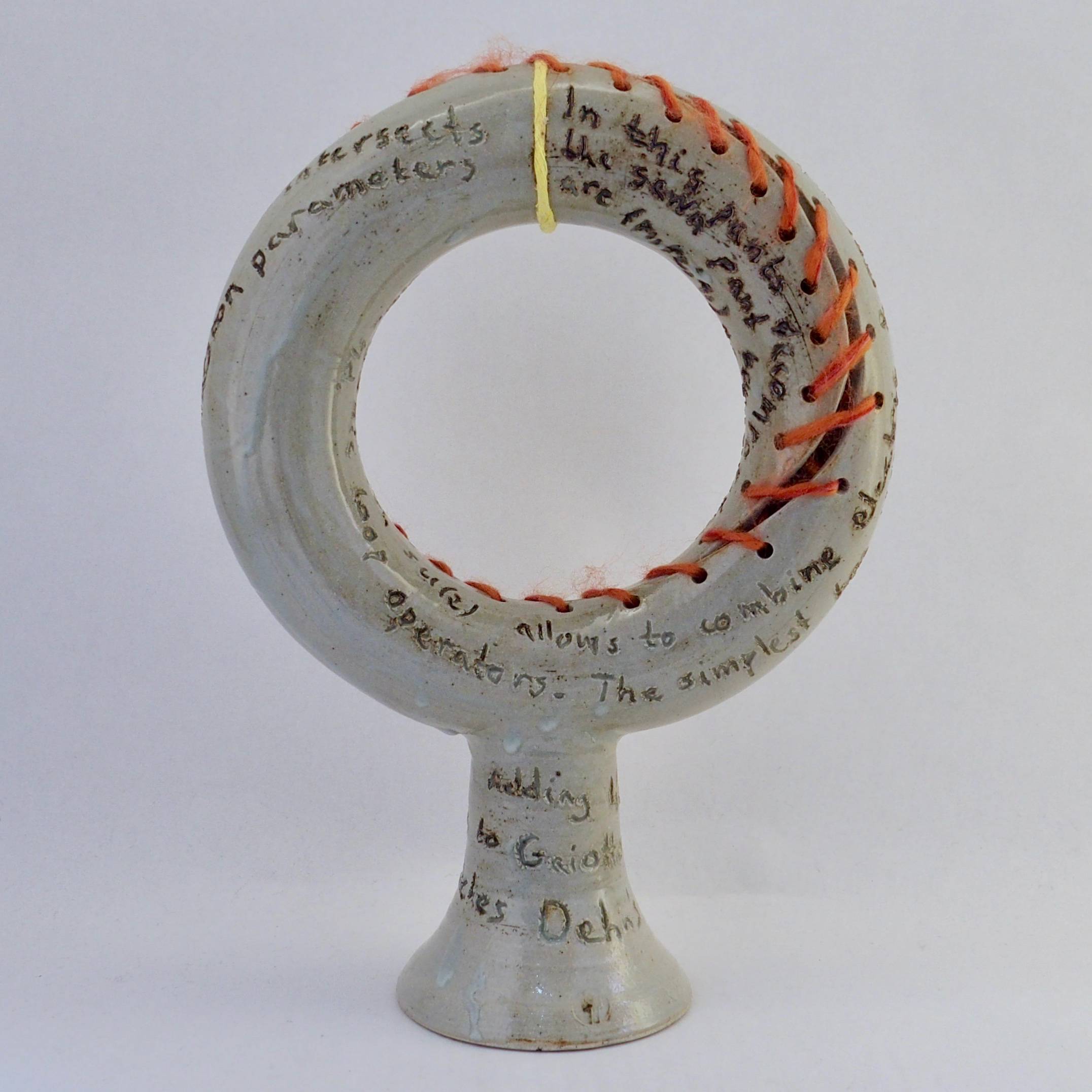}
%        	\subcaption{}
%        	\label{fig:2c}
\end{minipage}
 %add desired spacing between images, e. g., ~, \quad, \qquad, \hfill etc.	
\caption{``Sewing-1'', wheel thrown and altered stoneware, incised, iron oxide wash, 
celadon glaze and yarn. Triptych. Each $27\times20\times10$cm.}
\label{fig:sewing-1}
\end{figure}

The pieces in Figure~\ref{fig:cutting} are for illustration purposes. 
For my sculptures, I do cut them, but aim to keep the original shape of the torus. 
I also pierced holes along the cut before firing the pieces. After the works come out 
of the kiln, I use yarn or leather straps to ``sew'' the pairs of pants back into once-punctured tori. The resulting 
completed pieces form the triptych ``Sewing-1'', shown in Figure~\ref{fig:sewing-1}.

\noindent
The discussion so far concerned only the pants decomposition, not the curve on the surface. In 
all the works above I chose the curve to wrap the small cycle of the torus close to the top, represented 
by the yellow strings in Figure~\ref{fig:sewing-1}. 
As the classification outlined above assigns numbers with respect to a particular 
pants decomposition, the three different pieces have different Dehn-Thurston numbers describing 
the same curve. Those are $(p,q)=(0,1)$ for the image on the left $(1,0)$ in the middle and $(1,-1)$ on 
the right. The details of how to read the numbers are in references \cite{DMO, Dehn, Thurston, PH}.

The inscriptions on these and my other works outline the calculation of these parameters for 
the appropriate decomposition. It also includes details of the physics version of these parameters. 
As discussed towards the end of this manuscript, these are line operators for particles carrying 
electric and magnetic charges.

\section*{Four-Punctured Sphere}

\begin{figure}[h!tbp]
\centering
\begin{minipage}[b]{0.32\textwidth} 
	\includegraphics[width=\textwidth]{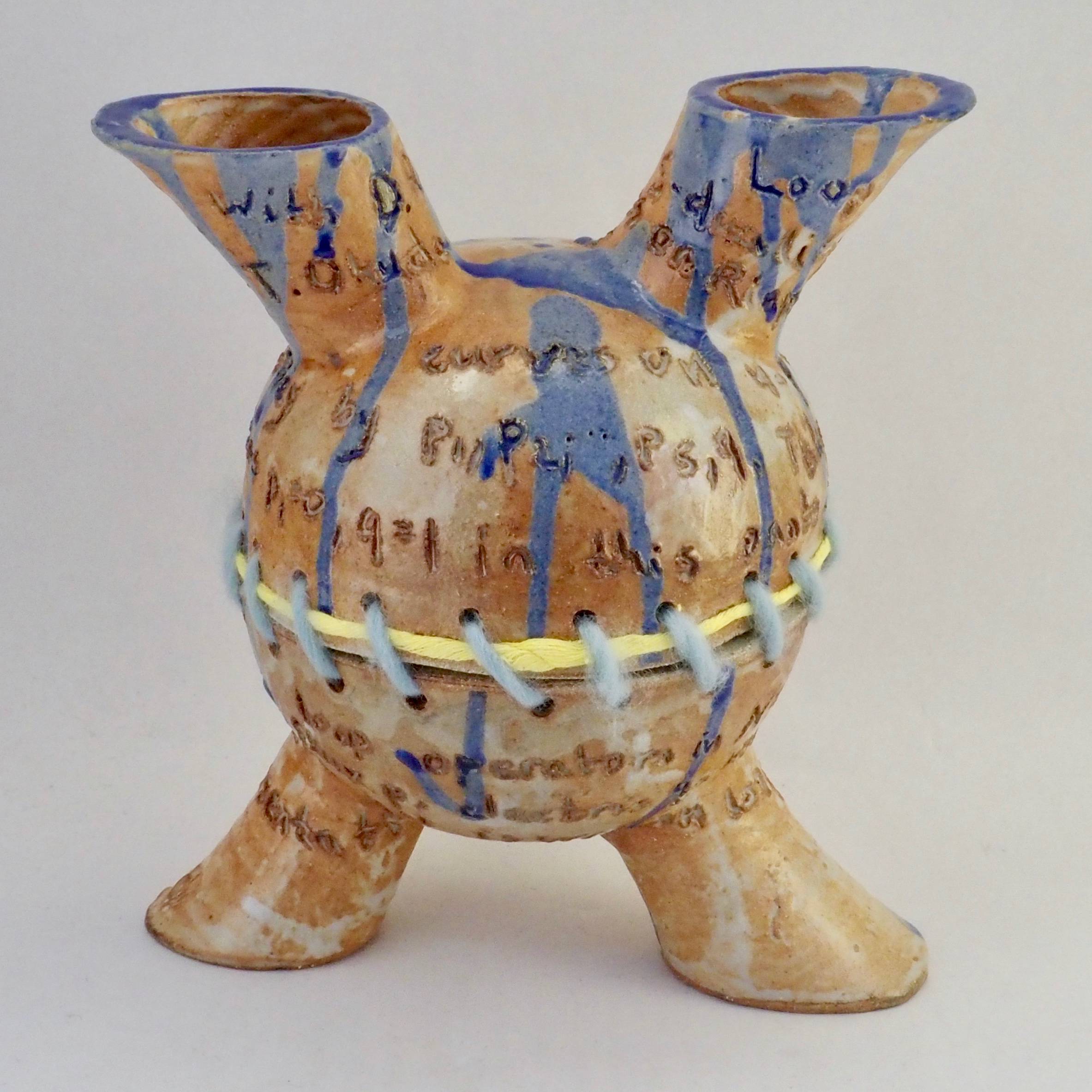}
        	\subcaption*{Horizontal cut.}
\end{minipage}
~ %add desired spacing between images, e. g., ~, \quad, \qquad, \hfill etc.	
\begin{minipage}[b]{0.32\textwidth} 
	\includegraphics[width=\textwidth]{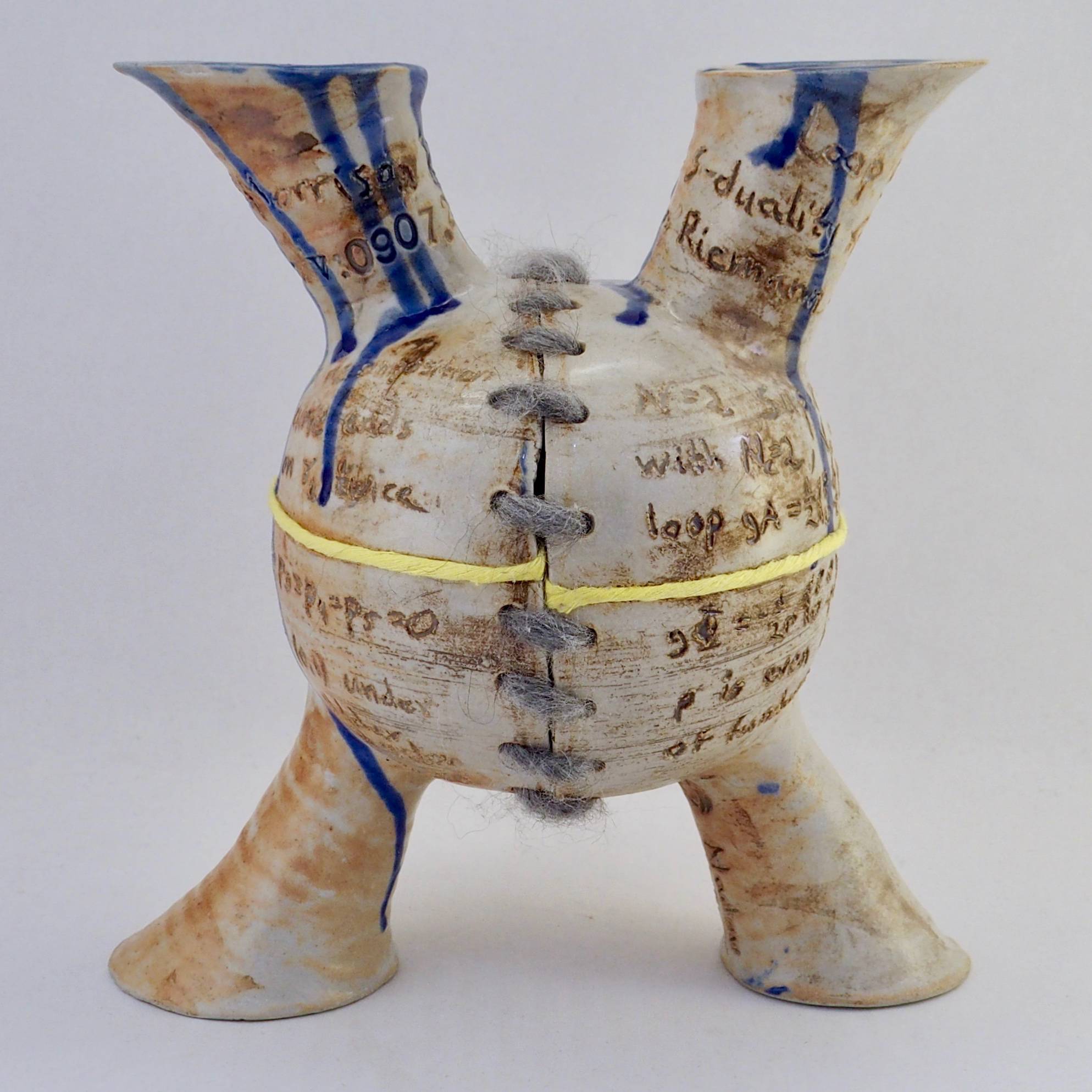}
        	\subcaption*{Vertical cut.}
\end{minipage}
~ %add desired spacing between images, e. g., ~, \quad, \qquad, \hfill etc.	
\begin{minipage}[b]{0.32\textwidth} 
	\includegraphics[width=\textwidth]{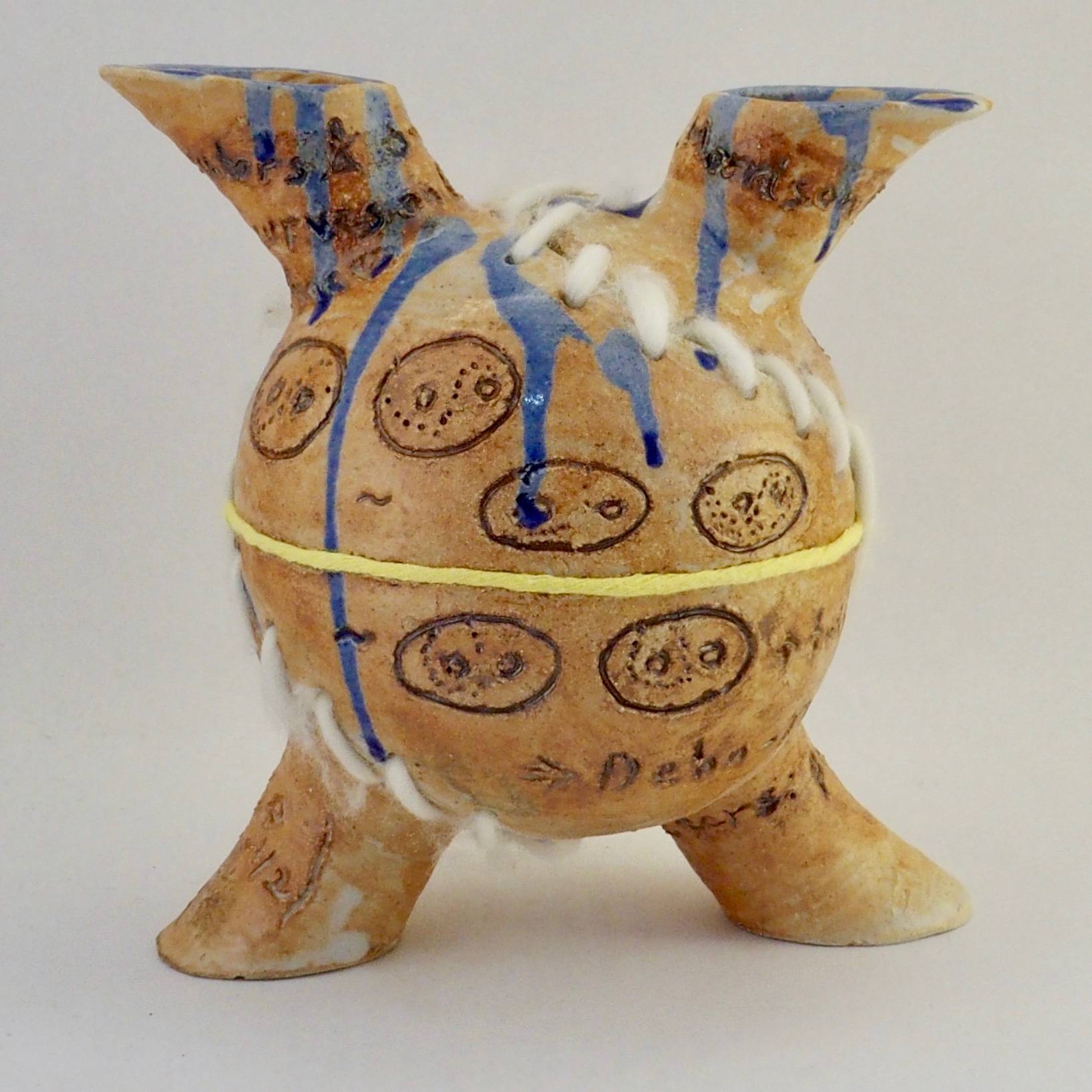}
        	\subcaption*{Diagonal cut.}
\end{minipage}
%add desired spacing between images, e. g., ~, \quad, \qquad, \hfill etc.	
\caption{Three pieces comprising the triptych ``Sewing-2''. Thrown and assembled stoneware, 
incised, iron oxide wash, shino and glue glaze and string. Each $20\times18\times13$cm.}
\label{fig:sewing-2}
\end{figure}

Our basic building block is the pair of pants, or the three-punctured sphere. Sewing two 
together eliminates two of the six punctures, resulting in the four-punctured sphere. 
It is easy to 
make a spherical object on the potter's wheel, and I attach to them 
four tubes (similar to the single leg above) to represent the punctures. 
The pieces are designed to stand on a pair of those tubes with the other pair
pointing upwards, see Figure~\ref{fig:sewing-2}.

As in the previous example, there is an infinite number of ways to decompose the four-punctured 
sphere into pairs of pants. Clay versions of three of them are shown 
in Figure~\ref{fig:sewing-2}. The three cuttings shown combine each pair of holes: The two 
bottom/top, the two right/left and the diagonals. Other cuttings connect the pairs along more 
complicated paths.

\noindent
The curve on the surface is represented again by the yellow string and as before, I chose the same curve 
on all three pieces, but the Dehn-Thurston parameters change, as they depend on the cutting. 
In the cutting on the left, parallel to the string, $p=0$ and $q=1$. In the middle cutting, the curve 
intersects the string at two points, and doesn't twist, so $p=2$ and $q=0$. In the right-most cutting, 
again it intersects at two points, so $p=2$, but the twist implies $q=1$.

In making these pieces I had to deal with their stability and the fact that ceramics get deformed 
during the drying and firing stages. I started by cutting the pieces while the clay was still wet 
and firing them as they should stand. I put heat resistant nichrome wire hooks to hold them in place and the 
two pieces on the right of Figure~\ref{fig:sewing-2} survived the firing, but one similar to that on 
the left and a further one deformed and were unusable. So the left one was fired as two separate 
pieces fired on other plates of clay that shrunk together with the hemispheres preserving their form.

\section*{Five-Punctured Sphere}
\begin{figure}[h!tbp]
	\centering
	\includegraphics[width=3.7in]{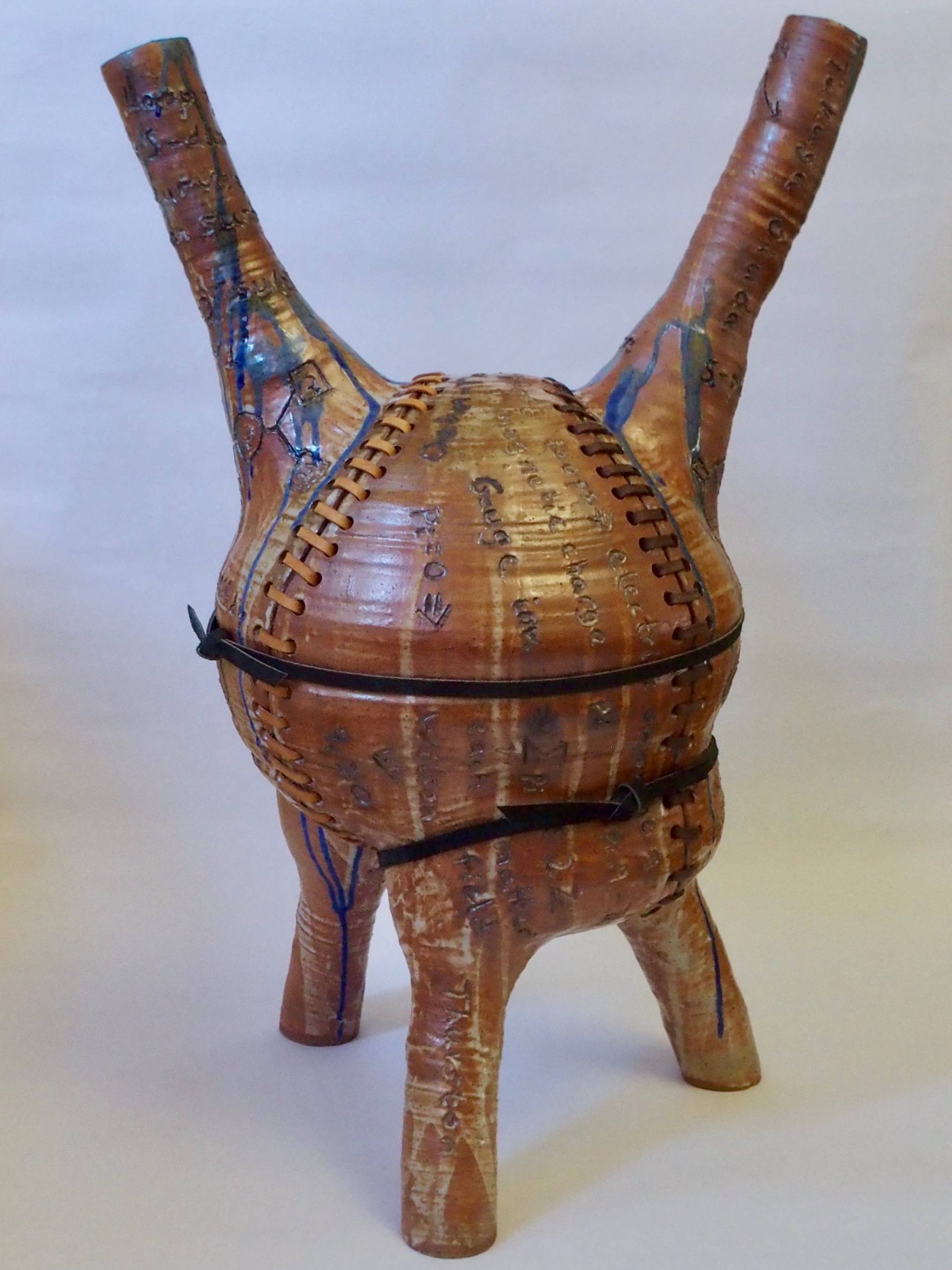}
	\caption{``Sewing-6''. Thrown and assembled stoneware with iron oxide wash, shino and blue 
	glazing and leather straps, $63\times42\times35$cm. The 5-punctured sphere is assembled 
	from three ``pairs of pants''.}
	\label{fig:sewing-6}
\end{figure}

A five-punctured sphere is divided into three pairs of pants along two different cuts. I have made 
two such pieces, one shown in Figure~\ref{fig:sewing-7} and one in Figure~\ref{fig:sewing-6}.

\begin{figure}[h!tbp]
\centering
\begin{minipage}[b]{0.4\textwidth} 
	\includegraphics[width=\textwidth]{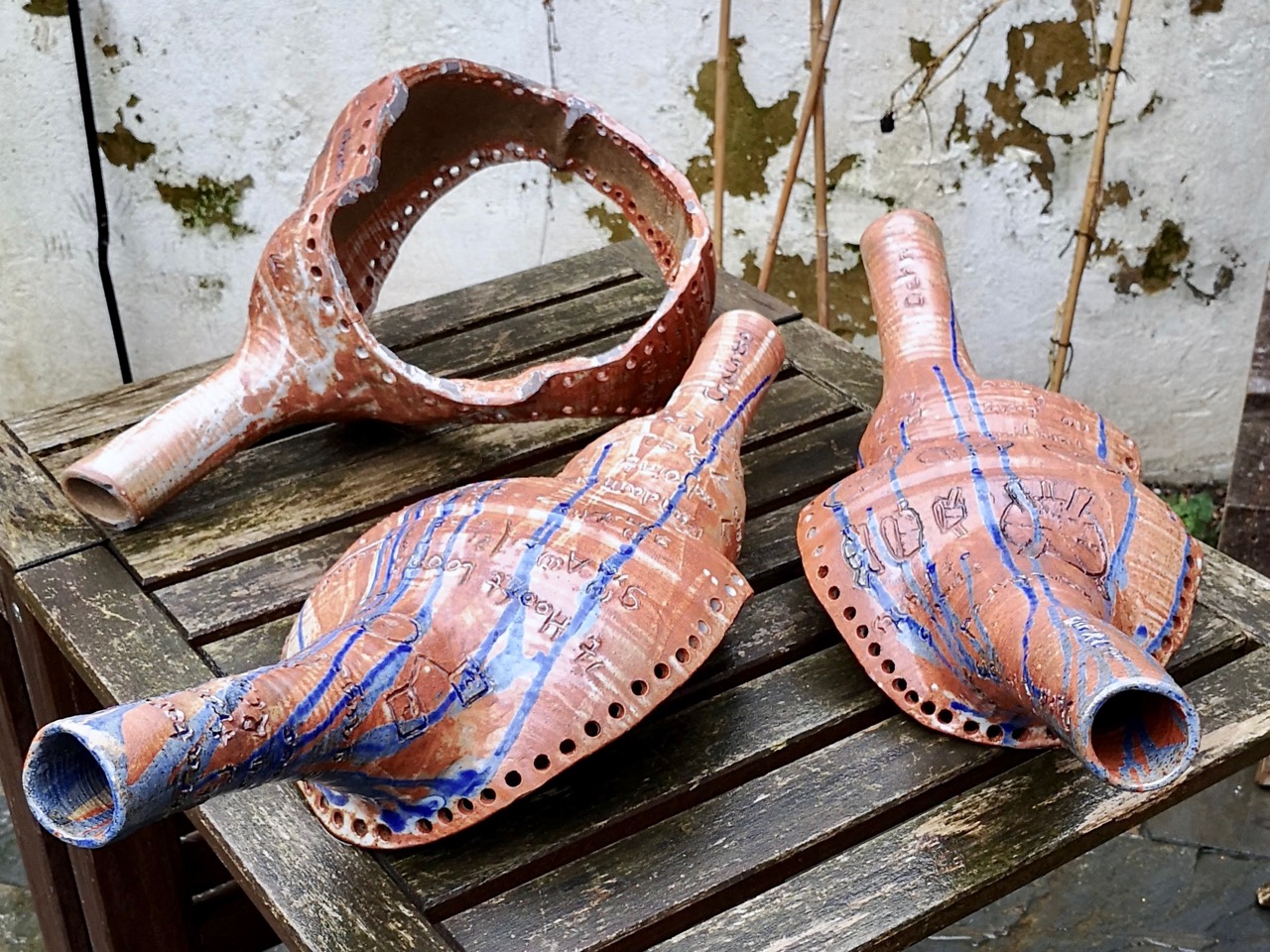}
        	\subcaption*{The three pairs of pants after being cut.}
\end{minipage}
~ %add desired spacing between images, e. g., ~, \quad, \qquad, \hfill etc.	
\begin{minipage}[b]{0.4\textwidth} 
	\includegraphics[width=\textwidth]{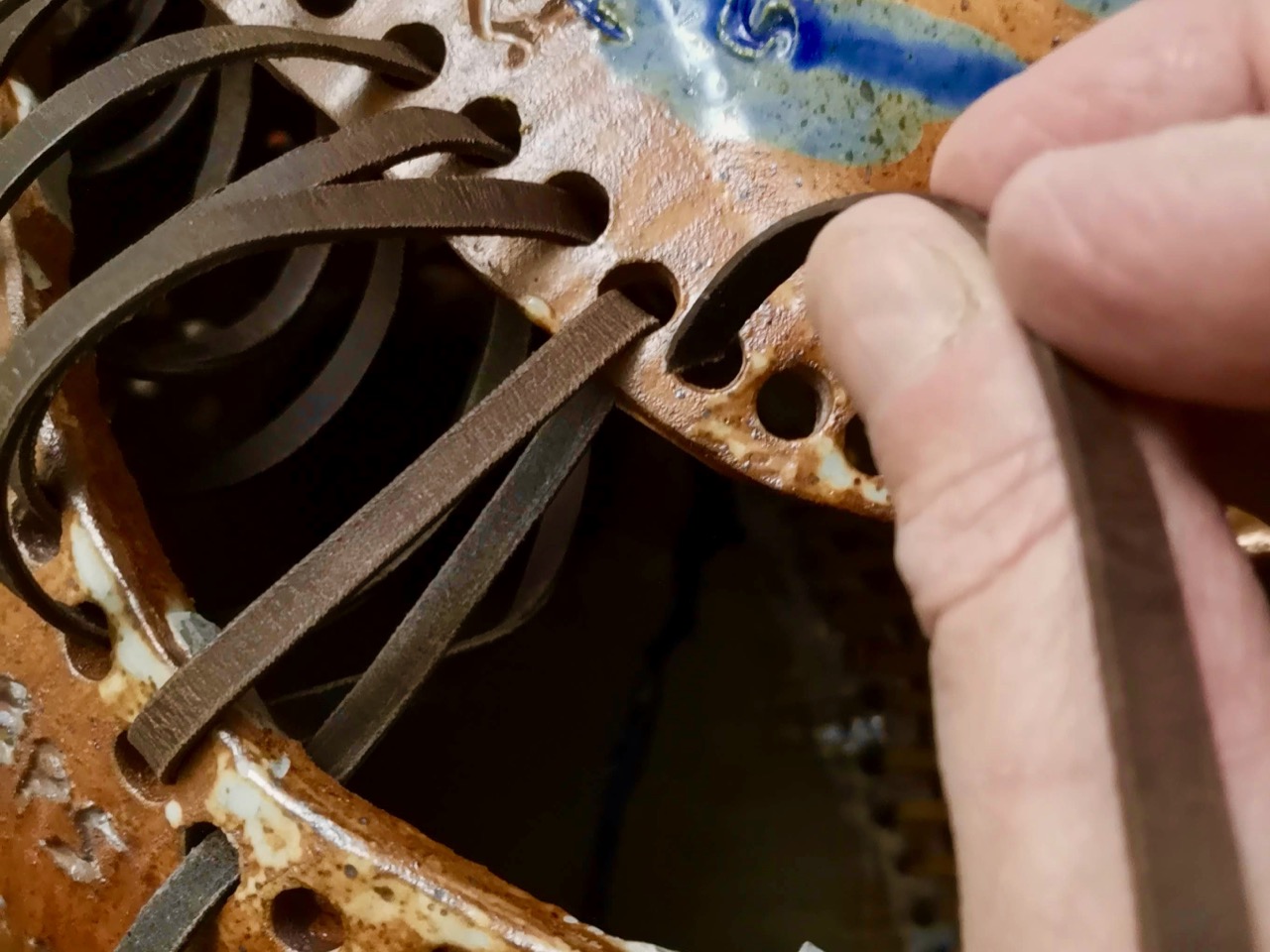}
        	\subcaption*{Sewing the pieces together with leather straps.}
\end{minipage}
\caption{Some steps in making the five-punctured sphere ``Sewing-6''.}
\label{fig:making}
\end{figure}

For stability of these pieces, I didn't fully cut them before the firing. Instead I prepared a perforated 
cut, leaving 1cm bridges every 5cm. The sewing holes were also pierced before the firing.

After the glaze firing, I used a diamond disc cutter to remove the bridges and separate the piece into 
the three pairs of pants. The results are shown at the left panel of Figure~\ref{fig:making}. I then 
reassembled them with leather straps, which is shown on the right panel. At the bottom left of that 
image, one can see on the inside of the seam an unglazed spot, which in fact is where one of the 
bridges was previously.

The resulting piece is shown in Figure~\ref{fig:sewing-6}. The non-selfintersecting curve, represented by 
the black leather straps is now composed of two disconnected pieces. One of its segments crosses 
both cuts and the other crosses only one. The Dehn-Thurston parameters are then $(p_1,q_1)=(4,0)$ 
for the right cut in the picture, crossed four times (only two seen) and $(p_2,q_2)=(2,0)$ for the 
left cut (only one seen).

\noindent
The Dehn-Thurston parameters for the analogous black straps in Figure~\ref{fig:sewing-7} are 
$(p_1,q_1)=(2,1)$ for the cut winding down and around the piece and $(p_2,q_2)=(0,1)$ for 
the lower-right cut.

\section*{Line Operators and Curves on Riemann Surfaces}

The real topic of my research \cite{DMO} is the classification of line operators in certain 
supersymmetric field theories. It would go beyond the scope of this article to explain it in detail, 
but for those somewhat versed in physics, these theories are generalizations of electromagnetism, 
and the line operators match the possible types of particles in these theories.

There is a mapping due to Gaiotto \cite{gaiotto} between these physical theories and 
Riemann surfaces, where alternative descriptions of the same underlying physical theory 
are related to different pants decompositions of the same surface. Our paper \cite{DMO} extended the mapping 
to objects in these theories (the particles or line operators) and curves on the Riemann surfaces.

According to Gaiotto's mapping, each cut on the surface corresponds to one copy of 
Maxwell theory (to be precise an $su(2)$ vector multiplet). These theories may have electrically 
charged particles (like an electron) and magnetically charged ones (which are not observed 
in nature, though there are active searches for them). 

\begin{figure}[h!tbp]
\centering
\begin{minipage}[b]{0.4\textwidth} 
	\includegraphics[width=\textwidth]{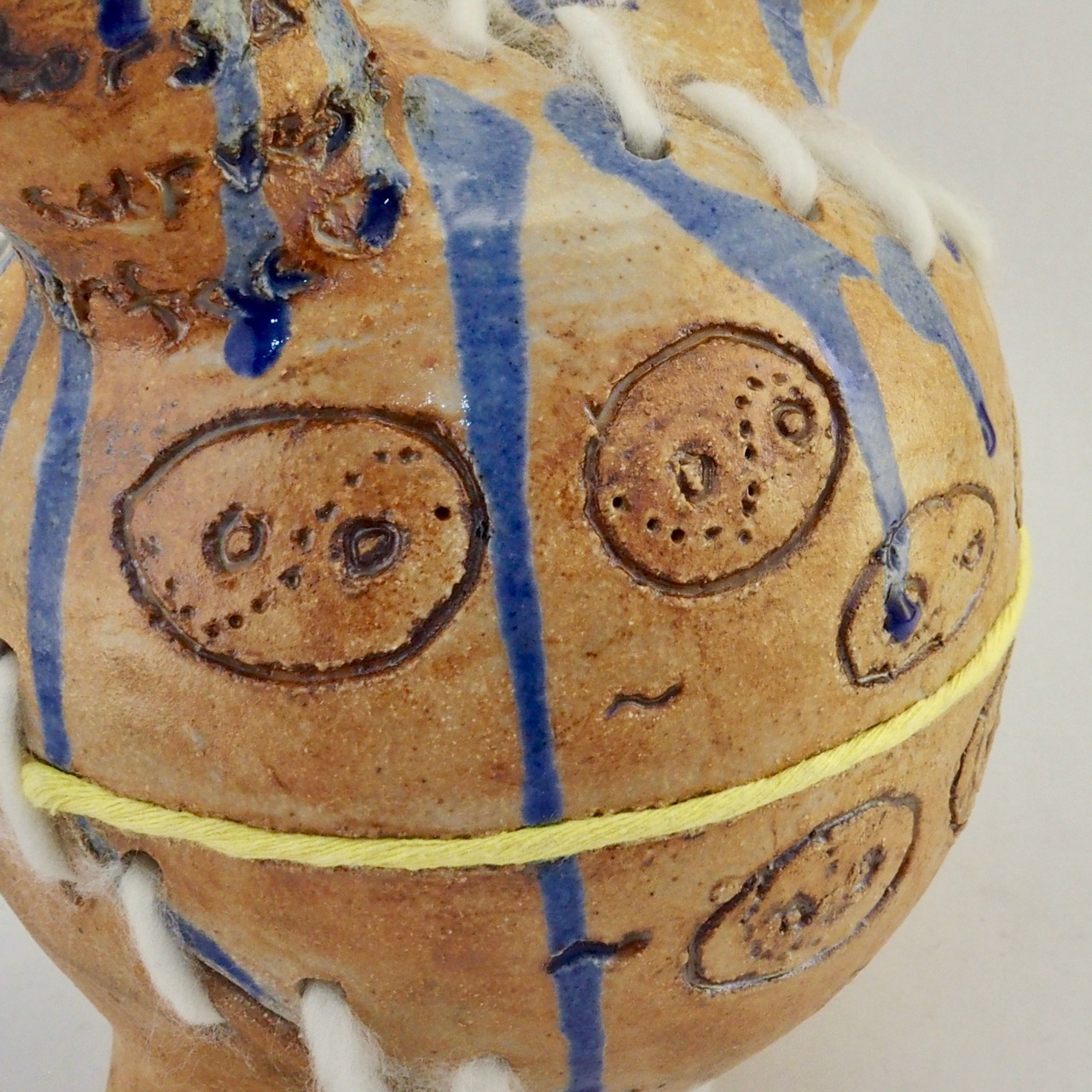}
%        	\subcaption{}
%        	\label{fig:details1}
\end{minipage}
~ %add desired spacing between images, e. g., ~, \quad, \qquad, \hfill etc.	
\begin{minipage}[b]{0.4\textwidth} 
	\includegraphics[width=\textwidth]{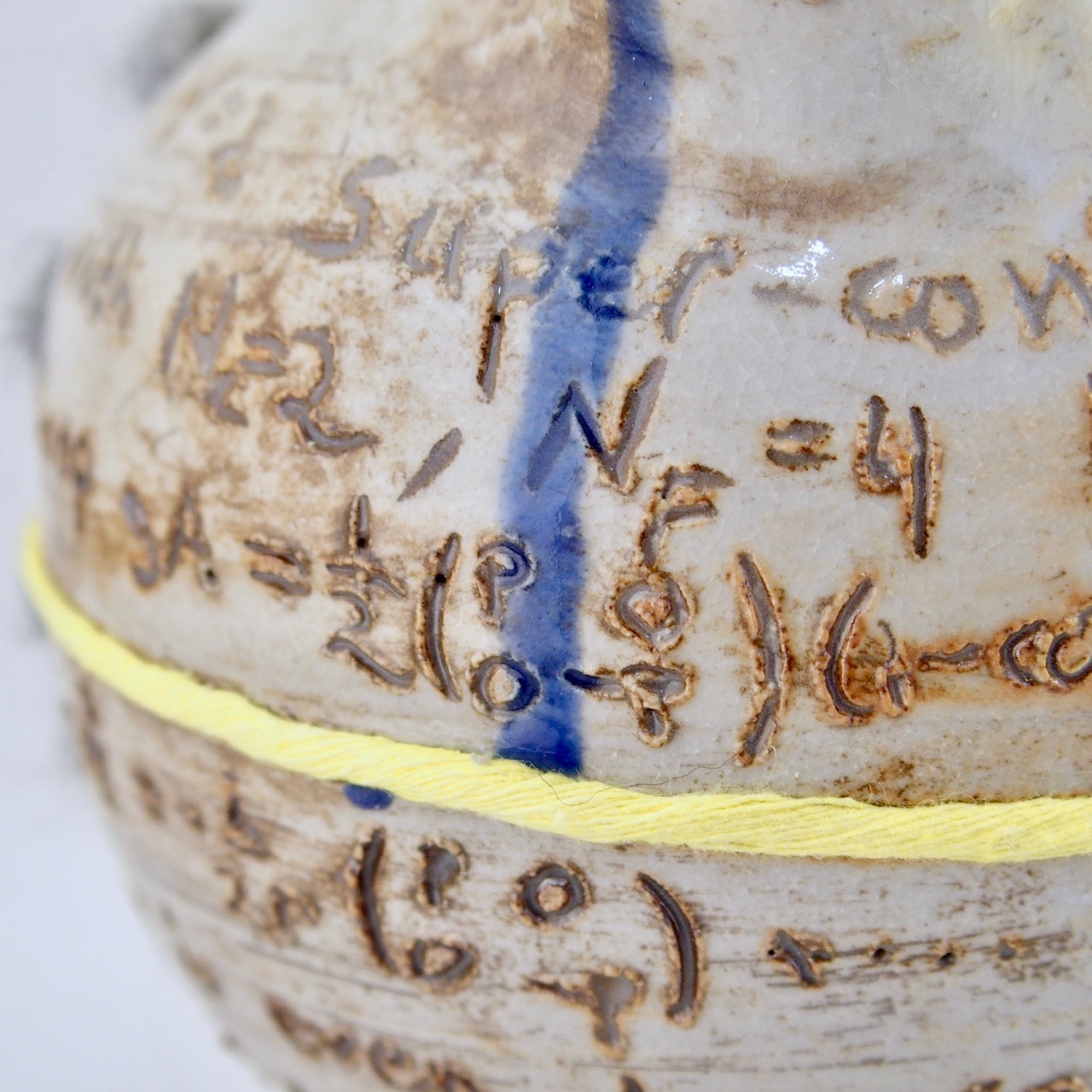}
%        	\subcaption{}
%        	\label{fig:details2}
\end{minipage}
\caption{Details from ``Sewing-2'': The ovals (left) are two pairs of pants with the 
dotted lines representing the yellow string. This is required for a proper calculation of 
the Dehn-Thurston parameters.
%\\
The physics avatar is also inscribed on the pieces (right). These are values of the gauge and scalar 
fields corresponding to a line operator with the 
same quantum numbers as the curve on the surface.}
\label{fig:details}
\end{figure}

\noindent
If both electrically and magnetically charged particles are allowed, so are particles carrying 
both charges (called dyons). There are certain conditions on the allowed charges 
carried by the particles (or lines), which we identified in our work. It turns out 
that the conditions exactly match those of the Dehn-Thurston parameters 
where $p_i$, the number of crossings, matches the magnetic charges and 
the twists $q_i$ match the electric charges.

Some details of the configurations associated to these line operators are 
shown on the right panel of Figure~\ref{fig:details}. On the left panel are 
details of calculating the Dehn-Thurston parameters, where the circle with 
two circles inside it is a flattened pair of pants. The dotted line indicates the 
path of the yellow string in this particular pants decomposition.

Another question in physics is how the electric and 
magnetic charges change in the different descriptions of the same theory. This 
was first studied in \cite{witten}, which supplied the answer in some cases. 
The full answer is given by our map to the curves on surfaces. The mapping of the 
Dehn-Thurston parameters for the same curve between different pants decompositions 
was studied in \cite{penner}, thus resolving this question.

%%%%%%%%%%%%%%%%%%%%%%%%%%%%%%%%%%%%%%%%%%%
\section*{Summary and Conclusions}
In this paper I presented a series of my ceramic artworks inspired by my research 
\cite{DMO}. I tried to combine aspects of the mathematics, how 
it drove the design of the 
artworks, some details of the ceramic processes and a little bit of the theoretical 
physics that was the actual focus of my original research. I plan to make more 
artworks in this series including higher genus surfaces.

These works are part of my outreach endeavors to present advanced mathematics and 
theoretical physics to a wider audience. I explained more about the general philosophy 
at Bridges-2019 in Linz \cite{bridges}.

%%%%%%%%%%%%%%%%%%%%%%%%%%%%%%%%%%%%%%%
\section*{Acknowledgements}

I thank the Racah Institute of the Hebrew University in Jerusalem for its hospitality during the writing of this 
paper.

%%%%%%%%%%%%%%%%%%%%%%%%%%%%%%%%%%%%%%%
% References %
    
{\setlength{\baselineskip}{13pt} % tighten line spacing for bibliography
\raggedright				% no right justification for References

} % end setlength, raggedright
   
\end{document}